\documentclass[a4paper]{article}

\usepackage{graphics}
\usepackage{geometry}
\usepackage{amssymb, amsmath}
\usepackage{epstopdf}
 \usepackage[applemac]{inputenc}

 \numberwithin{equation}{section}

 \newtheorem{theoreme}{Theorem}[section]
 \newenvironment{demo}{\textbf{Proof}}{$\square$}
 \newtheorem{lemme}[theoreme]{Lemma}
 
 \newtheorem{remarque}[theoreme]{Remark}
 \newtheorem{prop}[theoreme]{Proposition}
 
 \renewcommand{\Im}{\mathrm{Im}}
 \renewcommand{\Re}{\mathrm{Re}}
\date{}

\begin{document}

\title{On the theory of resonances in non-relativistic QED and related models}

\author{W. K. Abou Salem$^{a,\#}$, J. Faupin$^{b,*}$, J. Fr\"ohlich$^c$ and I. M. Sigal$^{a,\#}$}

\maketitle

\begin{abstract}
We study the mathematical theory of quantum resonances in the
standard model of non-relativistic QED and in Nelson's model. In
particular, we estimate the survival probability of metastable
states corresponding to quantum resonances and relate the resonances to poles of an
analytic continuation of matrix elements of the resolvent of the quantum Hamiltonian.
\end{abstract}

\section{Introduction}\label{sec:Introduction}

One of the early triumphs of Quantum Mechanics has been to enable one
to calculate the discrete energy spectrum and the corresponding
stationary states - eigenstates of the quantum Hamiltonian - of
atoms and molecules, neglecting their interactions with the
quantized electromagnetic field. However, if these interactions are
taken into account, stationary states corresponding to discrete
energies, save for the groundstate,  are absent. The data of atomic
and molecular spectroscopy can be interpreted in terms of the decay
of metastable states with energies close to the discrete energies,
or eigenvalues, of the non-interacting atoms or molecules. The decay
of these states is accompanied by emission of photons with nearly
discrete energies equal to differences between energies of
stationary states; (Bohr's frequency condition). These metastable
states are called ``(quantum) resonances''. Their analysis is the
subject of this paper: We further develop some key ingredients of
the mathematical theory of resonances for the standard model of
``non-relativistic quantum electrodynamics'' (QED) and for Nelson's
model of electrons interacting with quantized (longitudinal lattice)
vibrations, i.e., phonons. Due to the interactions of electrons with
massless field quanta - photons or phonons - the standard techniques
to analyze quantum-mechanical resonances developed during the past
thirty or so years (see, e.g., \cite{HS,{RS4}} and references
therein) cannot be applied to realistic models of atoms or
molecules. Our goal, in this paper, is to modify these techniques to
cover the present models.


Before introducing the models we explain the resonance problem in
general terms. Let $H_g$ be a quantum Hamiltonian, where $g$ is a
real parameter called the coupling constant. Assume there is a
one-parameter family of unitary transformations, $\theta \rightarrow
\mathcal{U}_{\theta},$ with $\theta \in \mathbb{R}$, s.t. the family
$H_{g,\theta} := \mathcal{U}_\theta H_g \mathcal{U}^{-1}_\theta $
has an analytic continuation in $\theta$ to a disc $D(0,\theta_0)$
in the complex plane. We call such an analytic continuation a
{\it complex deformation} of $H_g$. We note that, while the essential
spectrum  of $H_{g,\theta}$  usually changes dramatically under such
an analytic continuation, the eigenvalues are locally independent of
$\theta,$ for suitably chosen deformation transformations $\mathcal{U}_\theta$,
at least when they are isolated. Moreover, the real eigenvalues of
$H_{g,\theta}$ coincide with the eigenvalues of $H_g$.  The complex
eigenvalues of $H_{g,\theta},\ \Im \theta >0,$ are called the
(quantum $\mathcal{U}-$)resonance eigenvalues - or just resonance
eigenvalues - of the Hamiltonian $H_g$. The  transformations most
commonly used is the group of dilatations of positions and momenta
(see below), and the corresponding resonances are sometimes called
``dilatation resonances''.

It is plausible from our definition that resonances - at least for weakly coupled systems ($g$ small) - are closely related to eigenvalues of $H_{g=0}.$ But what is their physical significance?

Physically, one thinks of quantum resonances as long-lived metastable states or as ``bumps'' in the scattering
cross-section as a function of energy. The energies and life-times
of metastable states are given by the bumps' centers and the
inverse of the bumps' widths. A known approach to establish such
properties is as follows. Let ${\mathcal D}\subset {\mathcal H}$ denote the
dense linear subspace  of $\mathcal{U}-$entire vectors, i.e., vectors $\psi$
for which the family $\{\psi_{\theta}:=\mathcal{U}_{\theta}\psi\}_{\theta\in {\mathbb R}}$ has an analytic continuation to the entire complex plane. For such
vectors one has the ``Combes formula''
\begin{equation} \label{Combes}
( \psi,(H_g-z)^{-1}\psi ) =
(\psi_{\bar{\theta}},(H_{g,\theta}-z)^{-1}\psi_{\theta} ).
\end{equation}
If we continue the r.h.s. analytically, first in $\theta$ and
then in $z,$ then we see that matrix elements, $( \psi,
(H_g-z)^{-1}\psi ) ,$ of the resolvent, for $z\in {\mathbb C}, \ \
\Im z>0,$ and $\psi\in {\mathcal D},$ have an analytic continuation
in $z$ across the essential spectrum of $H_g$ to the ``second
Riemann sheet'' whenever the resolvent set of the operator
$H_{g,\theta},\ \Im \theta
>0,$ contains a part of  this essential spectrum\footnote{Here we use the terms Riemann sheet
and Riemann surface informally. However, we expect that the matrix
elements \eqref{Combes} do have a Riemann surface ramified at the
resonances of $H_g$.}. Clearly, eigenvalues of $H_{g,\theta},\ \Im
\theta
>0,$ in the lower complex half-plane, $\mathbb{C}^{-}$, are poles of
this analytic continuation, provided these eigenvalues are isolated.

In the latter case, the metastability property can be established (at least, for weakly coupled systems) by using
the relation - via the Fourier transform - between the propagator
and the resolvent, contour deformation and Cauchy's theorem (see
\cite{Hun, HS}). The ``bumpiness'' of the cross-section can be
connected to the resonance poles. The real and
imaginary parts of the resonance eigenvalues give the energy and the rate of decay, or the reciprocal life-time, of the metastable state.

The situation described above is exactly the one encountered in
Quantum Mechanics. In non-relativistic QED and phonon models, the
resonance eigenvalues are {\it not isolated}; more precisely, a
branch of essential spectrum is attached to every complex eigenvalue
of the deformed Hamiltonian $H_{g,\theta}$. This is due to the fact
that photons and phonons are {\it massless}. As a result, establishing the property of metastability  and the pole structure of the
resolvent (and the related bumpiness of the cross-section) becomes a challenge. In this paper, we prove, for non-relativistic QED and Nelson's model, the metastability
property of resonances and characterize them in terms of poles of a
meromorphic continuation  of the matrix elements of the resolvent
on a dense set of vectors.

Next, we introduce the models considered in this paper. The Hamiltonian of the QED model is defined as
\begin{equation}
\label{eq:NRQED} H_g^{SM} := \sum_{j=1}^N \frac{1}{2m_j} (p_j +
gA(x_j))^2 + V(x) + H_f,
\end{equation}
where $x=(x_1,\dots,x_N),$ $p_j = -i  \nabla_j$ denotes the momentum
of the $j^{th}$ particle and $m_j$ its mass, and $V(x)$ is the
potential energy of  the particle system. Furthermore, $A(y)$
denotes the quantized vector potential
\begin{equation}
\label{eq:VectorPot} A(y) =\sum_{\lambda \in \{-1, 1\}} \int
\frac{d^3k }{(2\pi)^3 } \frac{\chi(k)}{\sqrt{2|k|}} (e^{ik \cdot y}
\epsilon_{\lambda}(k)a_{\lambda}(k)+
e^{-ik \cdot y}\overline{\epsilon_{\lambda}(k)}a_{\lambda}^*(k)),
\end{equation}
where $k \in \mathbb{R}^3,$ $\chi$ is an ultraviolet cut-off that
vanishes sufficiently fast at infinity, and $\epsilon_{\lambda}(k), \lambda=-1,1,$ are two transverse polarization vectors, i.e., orthonormal vectors in
$\mathbb{R}^3\otimes \mathbb{C}$ satisfying $k \cdot \epsilon_{\lambda}(k) =0 ;$ moreover,
$H_f$ is the photon (quantized electromagnetic field) Hamiltonian
defined as
\begin{equation} \label{Hf}
H_f = \sum_{\lambda=-1,1}\int_{ \mathbb{R}^3 } \omega(k) a_\lambda^*( k ) a_\lambda( k ) dk,
\end{equation}
where $\omega(k) = |k|.$

The operator-valued distributions $a_{\lambda}(k)$ and $ a_{\lambda}^*(k)$ are annihilation and
creation operators acting on the symmetric Fock space
$\mathcal{F}_s$ over $L^2({\mathbb R}^3 \times {\mathbb Z}_2)$. They
obey the canonical commutation relations
\begin{equation} \label{commrelationsQED}
\left [ a^*_{\lambda}(k) , a^*_{\lambda'}(k') \right ] = \left [
a_{\lambda}(k) , a_{\lambda'}(k') \right ] = 0 \quad,\quad \left [
a_{\lambda}(k) , a_{\lambda'}^*(k') \right ] = \delta (k-k')\delta
_{\lambda, \lambda'},
\end{equation}
and
\begin{equation*}
a_\lambda (k)\Omega = 0,
\end{equation*}
where $\Omega\in \mathcal{F}_s$ is the vacuum vector.

The QED Hamiltonian $H_g^{SM}$ acts on the Hilbert space
$\mathcal{H}_{p} \otimes \mathcal{F}_s$, where $\mathcal{H}_{p}$ is
the Hilbert space for $N$ electrons, e.g. $\mathcal{H}_{p} =
\mathrm{L}^2(\mathbb{R}^{3N}),$ (neglecting permutation symmetry). In (\ref{eq:NRQED}), Zeeman terms
coupling the magnetic moments of the electrons to the magnetic field
are neglected.

Nelson's model describes non-relativistic particles without spin
interacting with a scalar, massless boson field. The Hamiltonian of the model acts on $\mathcal{H}_{p} \otimes \mathcal{F}_s$,
where $\mathcal{F}_s$ is the symmetric Fock space over
$\mathrm{L}^2(\mathbb{R}^3),$ and is given by
\begin{equation}\label{def_Hg}
H_g^N := H_{p} \otimes I + I \otimes H_f + W_g .
\end{equation}
Here, $H_{p} = \sum_{j=1}^N p_j^2/2m_j + V$ denotes an $N$-particle
Schr\"odinger operator on $\mathcal{H}_{p}$. We assume that its
spectrum, $\sigma( H_{p} ),$ consists of a sequence of discrete
eigenvalues, $\lambda_0, \lambda_1 ,\cdots ,$ below some real number
$\Sigma$ called the {\it ionization threshold}. 

For $k$ in $\mathbb{R}^3$, we denote by $a^*(k)$ and $a(k)$ the
usual phonon creation and annihilation operators on $\mathcal{F}_s$. They are operator-valued distributions obeying the canonical commutation relations
\begin{equation} \label{commrelations}
\left [ a^*(k) , a^*(k') \right ] = \left [ a(k) , a(k') \right ] =
0 \quad,\quad \left [ a(k) , a^*(k') \right ] = \delta (k-k').
\end{equation}
The operator associated with the energy of the free boson field,
$H_f$, is given by the expression \eqref{Hf}, except that the operators $a^*(k)$ and $a(k)$ now are scalar creation and
annihilation operators as given above. The interaction $W_g$ in $(\ref{def_Hg})$ is assumed to be of the form
\begin{equation}
W_g = g \phi ( G_x )
\end{equation}
where
\begin{equation}
\label{eq:FormFactor}
\phi ( G_x ) = \sum_{j=1}^N \int_{ \mathbb{R}^3 } \frac{ \chi(k) }{ |k|^{1/2-\mu} } \left [ e^{-ik \cdot x_j} a^*(k) + e^{ik \cdot x_j} a(k) \right ] dk.
\end{equation}
As above, the function $\chi(k)$ denotes an ultraviolet cut-off, and
the parameter $\mu$ is assumed to be positive. 

Next, we state our assumptions on the potential and the ultraviolet cut-off $\chi,$ in particular concerning analyticity under dilatations.

\begin{itemize}

\item[(A)]
The potential $V(x)$ is dilatation analytic, i.e., the
vector-function $\theta\mapsto V(e^\theta x) (-\Delta + 1)^{-1}$ has
an analytic continuation to a small complex disc $D(0,\theta_0)\subset \mathbb{C},$ for some $\theta_0>0.$

\end{itemize}

An example of a dilatation-analytic potential $V$ is the Coulomb
potential for $N$ electrons and one fixed nucleus located at
the origin. For a molecule in the Born-Oppenheimer approximation, the potential 
$V(x)$ is not dilatation-analytic. In this case, one has to use a
more general notion of distortion analyticity (see \cite{HS}), which
can be easily accommodated in our analysis.

\begin{itemize}
\item[(B)]
The function $\chi$ is dilatation analytic, i.e., $\theta \mapsto
\chi( e^{-\theta} k )$ has an analytic continuation from ${\mathbb
R}$ to the disc $D(0,\theta_0)$.

\end{itemize}

For instance, we can choose $\chi(k): = e^{-k^2 / \Lambda^2},$ for some fixed, arbitrarily large ultraviolet cut-off $\Lambda>0.$

Let $H_g$ denote either $H_g^{SM}$  or $H_g^{N}$. To
define quantum resonances for the Hamiltonian $H_g,$ we use  the
dilatations of electron positions  and of photon momenta:
$$x_j\rightarrow e^\theta x_j\  \mbox{and}\ k\rightarrow e^{-\theta} k,$$
where $\theta$ is a real parameter. Such dilatations are represented
by the one-parameter group of unitary operators,
$\mathcal{U}_{\theta},$ on the total Hilbert space ${\mathcal H}:=
{\mathcal H}_p \otimes {\mathcal F}_s$ of the system. This is one of the most important 
examples of the deformation groups mentioned above\footnote{See,
however, Remark \ref{transformations} on page 25.}. Following the
general prescription, we define, for $\theta\in {\mathbb R},$ the
family of unitarily equivalent Hamiltonians
\begin{equation} \label{I.10} H_{g,\theta} := \mathcal{U}_\theta H_g
\mathcal{U}^{-1}_\theta .
\end{equation}
By the above assumptions on $V$ and $\chi,$ the family
$H_{g,\theta}$ can be analytically extended, as a type-A family
in the sense of Kato, to all $\theta$ belonging to the disc
$D(0,\theta_0)$ in the complex plane, where $\theta_0$ is as in assumptions (A) and (B). The deformation resonances are
now defined as complex eigenvalues of $H_{g,\theta},\ \Im \theta
>0.$

Let  $\lambda_0 := \inf (\sigma (H_{g=0}))$. We consider the
eigenvalues $\lambda_{j}$ of $H_p$, or of $H_0:= H_{p}\otimes I +
I\otimes H_f,$ with $\lambda_{0} < \lambda_{j} < \Sigma.$   By the
renormalization group analysis in \cite{BFS1,BFS2,BFS3,Sigal}, we
know that, as the interaction between the non-relativistic particles
and the field is turned on, these eigenvalues  turn into resonances
$\lambda_{j,g},$ with $\Im\lambda_{j,g}<0 $ and these resonances are
$\theta-$independent; (see also \cite{F} for a somewhat different
model). Our goal is to investigate the properties of these
resonances, as described above.

To simplify our presentation, we assume that $\lambda_{j}$ is
non-degenerate, and we denote by $\Psi_j = \psi_j \otimes \Omega$ the
normalized, unperturbed eigenstate associated with $\lambda_{j}$. We
also assume that

\begin{itemize}
\item[(C)]
{\it Fermi's Golden Rule} (\cite{BFS1,BFS2,BFS3}) holds.
\end{itemize}

This condition implies that $\Im\lambda_{j,g}\le -\mathrm{c}_0
g^{2},$ for some positive constant $\mathrm{c}_0 ; $ see for example
\cite{BFS1,BFS2,BFS3}. 

The main results of this paper are summarized in the following theorems.

\begin{theoreme}\label{main_theorem}
Let $H_g$ be either $H_g^{SM}$  or $H_g^{N}$. Given $\Psi_j,$ and $\lambda_{j,g}$ as above, and under Assumptions (A)-(C)  formulated above, there exists some $g_0>0$
such that, for all $0<g < g_0$ and times  $t\ge 0,$
\begin{equation}
\label{eq:Main} \left ( \Psi_j , e^{-itH_g } \Psi_j \right ) = e^{-
i t \lambda_{j,g} } + O(g^{\alpha}),
\end{equation}
where $\alpha:= \frac{2+4\mu}{5+2\mu},$ with $\mu>0$ appearing in
(\ref{eq:FormFactor}) for the Nelson model, and $\alpha=\frac{2}{3}$ for QED.
\end{theoreme}

\begin{remarque}
We expect that our approach extends to situations where Fermi's Golden Rule condition fails, as long as $\Im\lambda_{j,g}<0$, and that we can improve the exponent of $g$ in the error term by using an initial state that is a better approximation of the ``resonance state''; see section 3.
\end{remarque}

\begin{remarque}
The analysis below, together with Theorem 3.3 in \cite{A-SF}, gives an adiabatic theorem for quantum resonances in non-relativistic QED.
\end{remarque}

Theorem \ref{main_theorem} estimates the survival probability,
$\left( \Psi_j , e^{-itH_g } \Psi_j \right )$, of the state
$\Psi_j$. Let $\gamma_{j,g}:=-\Im \lambda_{j,g}$ and $T_{j,g}:=1/
\gamma_{j,g}$. Theorem \ref{main_theorem} implies that
\begin{equation}
\| e^{-itH_g }\Psi_j - e^{- i t \lambda_{j,g} } \Psi_j \| = [1- e^{-
2t \gamma_{j,g} }+ O(g^{\alpha})]^{1/2},
\end{equation}
which is $\ll 1,$ for  $t \ll T_{j,g}$. This property is what we call
the {\it ``metastability''} of the resonance associated with the resonance
eigenvalue $\lambda_{j,g}$.

There is a dense linear subspace ${\mathcal D}\subset {\mathcal H}$
of vectors with the property that, for every $\psi\in {\mathcal D},$
the family $\{\mathcal{U}_{\theta}\psi\}_{\theta\in {\mathbb R}}$ of vectors
has an analytic extension in $\theta$ to the entire complex plane,
with $\mathcal{U}_{\theta}\psi\in {\mathcal D},$ for any $\theta\in {\mathbb
C}.$ Vectors in ${\mathcal D}$ are called dilatation-entire vectors.

Next, for $z_*\in {\mathbb C}$ and $0 \le \varphi_1 < \varphi_2 <
2\pi ,$
we define domains
$$W_{z_*}^{\varphi_1, \varphi_2} := \{ z\in {\mathbb C} | \ \  |z-z_*| < \frac{1}{2}
|\Im z_*|,\ \varphi_1 \le \arg(z-z_*) \le \varphi_2 \}.$$

Our second main result is the following theorem.

\begin{theoreme} \label{thm:resonpoles}
Let $H_g$ be either $H_g^{SM}$  or $H_g^{N}$. Let Conditions (A), 
(B) and (C) be satisfied, and let  $\lambda_{0} < \lambda_{j} < \Sigma$ be
an eigenvalue of $H_p$.
Then there are a constant $g_{*}>0$ and a dense set ${\mathcal D}'
\subset {\mathcal D}$ s.t., for $g < g_{*}$ and for all $\psi\in
{\mathcal D}',$ the function $$F_\psi (z):= ( \psi,(H_g -z)^{-1}\psi
)$$ has an analytic continuation in $z$
from the upper half-plane, across a neighbourhood of
$\lambda_j$, into the domain $W_{\lambda_{j,g}}^{\varphi_1,
\varphi_2}$, for some $\varphi_1 < \pi/2$ and $\varphi_2 > \pi$, and
this continuation satisfies the relations

\begin{equation}
\tag{i} F_\psi(z) = \frac{p(\psi)}{\lambda_{j,g}-z} + r(z; \psi),
\end{equation}
with
\begin{equation}
\tag{ii} |r(z; \psi)| \le \mathrm{C}(\psi)
|\lambda_{j,g}-z|^{-\beta},
\end{equation}
for some $\beta <1.$
Here $p(\psi)$ and $r(z; \psi)$ are
quadratic forms on the domain ${\mathcal D}'\times {\mathcal D}'$.
\end{theoreme}

\begin{remarque}
Since we can rotate the essential spectrum of $H_{g,\theta},
\theta\in D(0,\theta_0),$ in $\mathbb{C}^-$ using dilatation
analyticity, if $\theta_0<\frac{\pi}{2}$ is large enough, we expect
that $F_\psi(z)$ can be analytically continued in $z$ from the upper
half-plane into a neighbourhood of $\lambda_{j,g}$ that is larger
than $W_{\lambda_{j,g}}^{\varphi_1,\varphi_2}$ given in Theorem
\ref{thm:resonpoles}. In this case the quadratic form $r(w; \psi)$
would also depend on the homotopy class of the path along which
$F_\psi (z)$ is analytically continued from the upper half-plane to
the point $w$ in the vicinity of $\lambda_{j,g}.$
\end{remarque}

For an operator $A$ on the one-particle space $L^2(\mathbb{R}^3),$ we
denote by $d \Gamma ( A)$ its ``lifting'' to the Fock space
$\mathcal{F}_s,$ (second quantization). The set ${\mathcal D}'$ in Theorem \ref{thm:resonpoles} can be chosen explicitly as
$${\mathcal D}':= \{\psi \in {\mathcal D} | \ \ \| d \Gamma (
\omega^{-1/2}) (1 - P_{\Omega}) \psi \| < \infty \},$$ where $
P_{\Omega}$ is the projection onto the vacuum $\Omega$ in ${\mathcal
F}_s$,  for the Nelson model.  In this case  $\beta=
(1+\frac{2}{3}\mu)^{-1}.$ For QED, we define $${\mathcal
D}':= \{\psi \in {\mathcal D} | \ \ \|e^{\delta \langle x \rangle} d
\Gamma ( \omega^{-1/2}) (1 - P_{\Omega}) \psi \| < \infty\ \mbox{for
some}\ \delta >0 \}.$$ Since $\mathcal{U}_\theta d \Gamma (
\omega^{-1/2})= e^{\theta/2} d \Gamma (
\omega^{-1/2})\mathcal{U}_\theta$, the set ${\mathcal D}'$ is dense
in ${\mathcal D}$.


The main difficulty in the proofs of our main results comes from the fact that the
unperturbed eigenvalue $\lambda_{j}$ is the threshold of a branch of
continuous spectrum. To overcome this difficulty, we introduce an
infrared cut-off that opens a gap in the spectrum of $H_{g,\theta},$ and we control
the error introduced by opening such a gap using the fact that
the interaction between the electrons and the photons or phonons
vanishes sufficiently fast at low photon/phonon energies (see
\cite{BFS1,BFS2, BFP} and Eqn. \eqref{estimate_Wgleq} below).

Our paper is organized as follows. In Sections
\ref{sec:Hamiltonians}-\ref{sec:Ham} we prove Theorem
\ref{main_theorem} for the Nelson Hamiltonian, $H_g^{N}$. In Section
\ref{sec:QEDproof} we extend this proof to the QED Hamiltonian,
$H_g^{SM}$. Theorem \ref{thm:resonpoles} is proven in Section
\ref{sec:polesproof}.

As we were completing this paper, there appeared an e-print \cite{HHH} where lower and
upper bounds for the lifetime of the metastable states considered
in this paper are established by somewhat different techniques.

\vspace{0.5cm}

{\bf Acknowledgements}. J.Fr. and I.M.S. would like to thank M.
Griesemer for many useful discussions on related problems. J.Fa. is
grateful to I.M.S. and W.A.S. for hospitality at the University of
Toronto and I.M.S., and I.M.S. and W.A.S. are grateful to J.Fr. for hospitality at ETHZ. 

\section{Dilatation analyticity and IR cut-off Hamiltonians}\label{sec:Hamiltonians}

Let $H_g = H_g^N$ be the Hamiltonian defined in $(\ref{def_Hg})$. We
begin this section with a discussion of the dilatation deformation 
$H_{g,\theta}$ of $H_g$ defined in the introduction, Eqn
\eqref{I.10}. As was already mentioned above, by the above
assumptions on $V$ and $\chi,$ the family $H_{g,\theta}$ can be
analytically extended to all $\theta$ belonging to a disc
$D(0,\theta_0)$ in the complex plane. The relation
$H_{g,\theta}^* = H_{g,\overline{\theta}}$
holds for real $\theta$ and extends by analyticity to $\theta\in
D(0,\theta_0).$ A direct computation gives
\begin{equation*}
H_{g,\theta} =  H_{p,\theta} \otimes I + e^{-\theta} I \otimes H_f +
W_{g,\theta},
\end{equation*}
where $H_{p,\theta} = \mathcal{U}_\theta H_{p}
\mathcal{U}_\theta^{-1}$ and $W_{g,\theta} := \mathcal{U}_\theta W_g
\mathcal{U}_\theta^{-1}$. Note that $W_{g,\theta} = g \phi(
G_{x,\theta} )$, with
\begin{equation}\label{def_Gxtheta}
G_{x,\theta} (k) = e^{-(1+\mu)\theta} \frac{ \chi ( e^{-\theta} k )
}{ |k|^{1/2-\mu} } e^{-ik.x}.
\end{equation}

We now introduce an infra-red cut-off Hamiltonian
\begin{equation}\label{decomposition_Hg}
H_{g,\theta}^\sigma := H_{p,\theta} \otimes I + e^{-\theta} I
\otimes H_f + W_{g,\theta}^{ \geqslant \sigma },
\end{equation}
where $W_{g,\theta}^{ \geqslant \sigma } := g \phi(
G_{x,\theta}^{ \leqslant \sigma })$, and $G_{x,\theta}^{ \leqslant \sigma } := \kappa_\sigma
G_{x,\theta}$. Here $\kappa_\sigma$ is an infrared cut-off
function that we can choose, for instance, as $\kappa_\sigma =
\mathbf{1}_{ |k| \geq \sigma }$. We also define
\begin{equation}
W_{g,\theta}^{ \leqslant \sigma } := W_{g,\theta} - W_{g,\theta}^{
\geqslant \sigma } = g \phi ( G_{x,\theta}^{ \geqslant \sigma } ),
\end{equation}
where $G_{x,\theta}^{ \geqslant \sigma } := (1-\kappa_\sigma) G_{x,\theta}$. We then have that
\begin{equation}\label{2.6}
H_{g,\theta}= H_{g,\theta}^\sigma + W_{g,\theta}^{\leqslant\sigma}.
\end{equation}

We denote by $\mathcal{F}_s^{ \geqslant \sigma }$ and
$\mathcal{F}_s^{ \leqslant \sigma }$ the symmetric Fock spaces over
$\mathrm{L}^2( \{ k \in \mathbb{R}^3 : |k| \geq \sigma \} )$ and
$\mathrm{L}^2( \{ k \in \mathbb{R}^3 : |k| \leq \sigma \} ),$
respectively. It is well-known that there exists a unitary operator
$\mathcal{V}$ that maps $\mathrm{L}^2(\mathbb{R}^{3N} ;
\mathcal{F}_s)$ to $\mathrm{L}^2(\mathbb{R}^{3N} ; \mathcal{F}_s^{
\geqslant \sigma }) \otimes \mathcal{F}_s^{ \leqslant \sigma }$, so
that
\begin{equation}\label{isomorphism_Hgsigma}
\mathcal{V} H_{g,\theta}^\sigma \mathcal{V}^{-1} = H_{g,\theta}^{
\geqslant \sigma } \otimes I + e^{-\theta} I \otimes H_f^{ \leqslant
\sigma }.
\end{equation}
Here, $H_{g,\theta}^{ \geqslant \sigma }$ acts on
$\mathrm{L}^2(\mathbb{R}^{3N} ; \mathcal{F}_s^{ \geqslant \sigma })$
and is defined by
\begin{equation}
H_{g,\theta}^{ \geqslant \sigma } := H_{p,\theta} +e^{-\theta}
H_f^{ \geqslant \sigma } + W_{g,\theta}^{ \geqslant \sigma }.
\end{equation}
The operators $H_f^{ \geqslant \sigma }$ and $H_f^{ \leqslant \sigma
}$ denote the restrictions of $H_f$ to $\mathcal{F}_s^{ \geqslant
\sigma }$ and $\mathcal{F}_s^{ \leqslant \sigma }$ respectively. The unitary operator $\mathcal{V}$ will be sometimes dropped in the sequel if no confusion may arise. We
note the following estimate that will often be used in this paper:
\begin{equation}\label{estimate_Wgleq}
\left \|  W_{g,\theta}^{ \leqslant \sigma } \left [ H_f + 1 \right
]^{-1} \right \| \leq \mathrm{C} g \sigma^{1/2 + \mu},
\end{equation}
where $\mu > 0, \ \ \mathrm{C}$ is a positive constant, and
$\theta\in D(0,\theta_0).$
\newline

We now consider an unperturbed isolated eigenvalue $\lambda_{j}$ of
$H_0.$ To simplify our analysis, we assume that $\lambda_{j}$ is
non-degenerate. Let
\begin{equation}
\mathrm{d}_j := \mathrm{dist}( \lambda_{j} ; \sigma( H_{p} )
\backslash \{ \lambda_{j} \} ),
\end{equation}
which is positive. It is shown in \cite{BFS1,BFS2,BCFS} that, as the perturbation $W_g$
is turned on, the eigenvalue $\lambda_{j}$ turns into a resonance
$\lambda_{j,g}$ of $H_g.$ In other words, for $\theta\in
D(0,\theta_0)$ with $\mathrm{Im}( \theta )>0,$ there exists a
non-degenerate eigenvalue $\lambda_{j,g}$ of $H_{g,\theta}$ {\it
not} depending on $\theta,$ with $\Re\lambda_{j,g} = \lambda_{j}+
O(g^2)$, $\mathrm{Im} \lambda_{j,g} =O(g^2)$, and, if
 Fermi's Golden Rule condition holds, $\mathrm{Im}\lambda_{j,g}\le
-\mathrm{c}_0 g^{2},$ for some positive constant $\mathrm{c}_0.$
Similarly, the operator $H_{g,\theta}^{ \geqslant \sigma }$ has an
eigenvalue $\lambda_{j,g}^{ \geqslant \sigma }$ bifurcating from the
eigenvalue $\lambda_j$ of $H_0$ having the same properties as
$\lambda_{j,g},$ with the important exception that
$\lambda_{j,g}^{\geqslant \sigma}$ depends on $\theta.$ The reason
for this is that $H_{g,\theta +r}^{ \geqslant \sigma }\ne {\mathcal
U}_{r}H_{g,\theta}^{ \geqslant \sigma }{\mathcal U}_{-r}, \ \ r\in
{\mathbb R} .$ Furthermore, we have the crucial property (see
Proposition \ref{theorem_Hgeqsigma}) that the eigenvalue
$\lambda_{j,g}^{ \geqslant \sigma }$  of  $H_{g,\theta}^{ \geqslant
\sigma }$ is isolated from the rest of the spectrum of
$H_{g,\theta}^{ \geqslant \sigma }$. More precisely,
\begin{equation}
\mathrm{dist} \left ( \lambda_{j,g}^{ \geqslant \sigma } , \sigma (
H_{g,\theta}^{ \geqslant \sigma } ) \setminus \{ \lambda_{j,g}^{
\geqslant \sigma } \} \right ) \ge \mathrm{C} \sigma ,
\end{equation}
for some positive constant $\mathrm{C}$ independent of $\sigma.$

It is tempting to treat $H_{g,\theta}$ as a perturbation of
$H_{g,\theta}^{ \geqslant \sigma }$. However, we have to take care
of the difference between $\lambda_{j,g}$ and $\lambda_{j,g}^{
\geqslant \sigma }$. In order to deal with this problem, we
``renormalize'' the unperturbed part $H_{g,\theta}^{ \sigma }$ by
setting
\begin{equation}\label{def_tildeHg}
\widetilde{H}_{g,\theta}^{ \sigma } = H_{g,\theta}^{ \sigma } +
\left ( \lambda_{j,g} - \lambda_{j,g}^{ \geqslant \sigma } \right )
\mathcal{V}^{-1} (P_{g,\theta}^{ \geqslant \sigma } \otimes I)
\mathcal{V}.
\end{equation}
Here $P_{g,\theta}^{ \geqslant \sigma }$ denotes the spectral
projection onto the eigenspace associated with the eigenvalue
$\lambda_{j,g}^{ \geqslant \sigma }$ of $H_{g,\theta}^{ \geqslant
\sigma }$. As in $(\ref{isomorphism_Hgsigma})$, we have the
representation
\begin{equation}
\mathcal{V} \widetilde{H}_{g,\theta}^{ \sigma } \mathcal{V}^{-1} =
\widetilde{H}_{g,\theta}^{ \geqslant \sigma } \otimes I +
e^{-\theta} I \otimes H_f^{ \leqslant \sigma },
\end{equation}
where we have set
\begin{equation}\label{def_tildeHg_2}
\widetilde{H}_{g,\theta}^{ \geqslant \sigma } = H_{g,\theta}^{
\geqslant \sigma } + \left ( \lambda_{j,g} - \lambda_{j,g}^{
\geqslant \sigma } \right ) P_{g,\theta}^{ \geqslant \sigma }.
\end{equation}
By $(\ref{def_tildeHg_2})$, we see that $\lambda_{j,g}$ is a
non-degenerate eigenvalue of $\widetilde{H}_{g,\theta}^{ \geqslant
\sigma }$. In Proposition \ref{lambda_diff_estimate} we will show
that there exists a
positive constant $\mathrm{C}$ such that
\begin{equation}
\left | \lambda_{j,g} - \lambda_{j,g}^{ \geqslant \sigma } \right |
\leq \mathrm{C} g^2 \sigma^{1+\mu},
\end{equation}
and that the operator $\widetilde{H}_{g,\theta}^{ \geqslant \sigma
}$ still has a gap of order $O(\sigma)$ around $\lambda_{j,g}.$ Then
the decomposition $(\ref{decomposition_Hg})$ is replaced by
\begin{equation} \label{2.16}
H_{g,\theta} = \widetilde{H}_{g,\theta}^\sigma +
\widetilde{W}_{g,\theta}^{ \leqslant \sigma } ,
\end{equation}
where
\begin{equation}
\label{eq:tildeW}
\widetilde{W}_{g,\theta}^{\leqslant\sigma}=W_{g,\theta}^{ \leqslant
\sigma } - \left ( \lambda_{j,g} - \lambda_{j,g}^{ \geqslant \sigma
} \right ) \mathcal{V}^{-1} P_{g,\theta}^{ \geqslant \sigma }
\otimes I \mathcal{V} .
\end{equation}
Let $H_\star^\#$ denote one of the operators $H_g$, $H_{g,\theta}$,
$H_{g,\theta}^{ \sigma }$ or $H_{g,\theta}^{ \geqslant \sigma }$. We
write its resolvent by using the notation $R_\star^\#(z) = \left [
H_\star^\# -z \right ]^{-1}$. Similarly, we define
$\widetilde{R}_\star^\#(z) = \left [ \widetilde{H}_\star^\# -z\right
]^{-1}$.
\newline

\section{Proof of Theorem \ref{main_theorem}}\label{sec:MainResult}

We begin with some notation. We consider an interval $\mathrm{I}$ of
size $\delta,$ containing $\lambda_{j},$ such that
$\delta<\frac{1}{2} \mathrm{d}_j.$  For concreteness, let
\begin{equation}
\mathrm{I} = \left ( \lambda_{j} - \frac{\delta}{2} , \lambda_{j} +
\frac{\delta}{2} \right ).
\end{equation}
Define, in addition,
\begin{equation}
\mathrm{I}_1 = \left ( \lambda_{j} - \frac{\delta}{4} , \lambda_{j}
+ \frac{\delta}{4} \right ).
\end{equation}
We consider a smooth function $f \in \mathrm{C}_0^\infty (
\mathrm{I} ), \ \ \mathrm{Ran}(f)\in [0,1],$ such that $f=1$ on
$\mathrm{I}_1$. It is known that there exists an almost analytic
extension $\widetilde{f}$ of $f$ such that
\begin{equation}
\widetilde{f} = 1 \text{ on } \left \{ z \in \mathbb{C}| \ \
\mathrm{Re}(z) \in \mathrm{I}_1 \right \} \quad,\quad \mathrm{supp}
( \widetilde{f} ) \subset \left \{ z \in \mathbb{C} | \ \
\mathrm{Re}(z) \in \mathrm{I} \right \},
\end{equation}
and $\left | ( \partial_{ \bar{z} } \widetilde{f} ) (z) \right | =
O( \delta^{-1}| \mathrm{Im}(z)/\delta |^n ),$ for any $n\in {\mathbb
N}.$ We shall use these properties of $\widetilde{f}$ in the sequel.

We begin with the following proposition.
\begin{prop}\label{pr:main}
Given $H_g,$ $\Psi_j,$ $\lambda_{j,g}$ and $f$ as above, there
exists $g_0>0$ such that, for all $0<g \le g_0,$ $\delta =
\mathrm{C}\sigma$, $\mathrm{C}>1,$ and $\sigma = g^{
2-\frac{2+4\mu}{5+2\mu} },$
\begin{equation}
\left ( \Psi_j , e^{-itH_g } f(H_g)\Psi_j \right ) = e^{- i t
\lambda_{j,g} } + O(g^{
\frac{2+4\mu}{5+2\mu} }),
\end{equation}
for all times $t\ge 0.$
\end{prop}

We divide the proof of Proposition \ref{pr:main} into several steps,
deferring the proof of some   technical ingredients to the following
section. We extend a method due to Hunziker to prove Proposition
\ref{pr:main}, see \cite{Hun} or \cite{HS}. Let ${\mathcal
N}(\theta)$ be a punctured neighbourhood of $\lambda_{j}$ such that
${\mathcal N}(\theta)\cap \sigma(\widetilde{H}_{g,\theta}^{
\geqslant \sigma }) = \lambda_{j,g}$ and $\mathrm{I} \subset
{\mathcal N}(\theta)\cup\{\lambda_{j}\}.$ Let $\Gamma\subset
{\mathcal N}(\theta)$ be a contour that encloses I and
$\lambda_{j,g}.$ For $z$ inside $\Gamma,$ we have that
\begin{equation}\label{decomp_R_g,theta}
\widetilde{R}_{g,\theta}^{ \geqslant \sigma } (z) = \frac{
P_{g,\theta}^{ \geqslant \sigma } }{ \lambda_{j,g} - z } +
\widehat{R}_{g,\theta}^{ \geqslant \sigma } (z),
\end{equation}
where $P_{g,\theta}^{ \geqslant \sigma }$ denotes the spectral
projection onto the eigenspace associated to the eigenvalue
$\lambda_{j,g}$ of $\widetilde{H}_{g,\theta}^{ \geqslant \sigma }$,
that is
\begin{equation}
P_{g,\theta}^{ \geqslant \sigma } = \frac{1}{2\pi i}
\oint_{\mathcal{C}} \widetilde{R}_{g,\theta}^{ \geqslant \sigma }
(z) dz,
\end{equation}
where $\mathcal{C}$ denotes a circle centered at $\lambda_{j,g}$
with radius chosen so that $\mathcal{C} \subset \rho( H_{g,\theta}^{
\geqslant \sigma } )\cap {\mathcal N}(\theta),$ and the {\it regular
part}, $\widehat{R}_{g,\theta}^{ \geqslant \sigma } (z),$ is given
by
\begin{equation}
\label{3.7}
\widehat{R}_{g,\theta}^{ \geqslant \sigma } (z) :=
\widetilde{R}_{g,\theta}^{ \geqslant \sigma } (z)(1-P_{g,\theta}^{
\geqslant \sigma })= R_{g,\theta}^{ \geqslant \sigma }
(z)(1-P_{g,\theta}^{ \geqslant \sigma }),
\end{equation} which can be written as
\begin{equation}
\label{eq:Regular} \widehat{R}_{g,\theta}^{ \geqslant \sigma } (z) =
\frac{1}{2\pi i} \oint_{\Gamma} \widetilde{R}_{g,\theta}^{ \geqslant
\sigma } (w)(w-z)^{-1} dw,
\end{equation}
where $z$ is inside $\Gamma.$ Note that
\begin{equation}
\label{eq:SpectralDecomp} \widehat{R}^{\geqslant\sigma}_{g,\theta}
P_{g,\theta}^{\geqslant\sigma} = P_{g,\theta}^{\geqslant\sigma}
\widehat{R}^{\geqslant\sigma}_{g,\theta} = 0,
\end{equation}
and
\begin{equation}
\label{eq:SpectralProj} (P_{g,\theta}^{ \geqslant \sigma })^2 =
P_{g,\theta}^{ \geqslant \sigma }.
\end{equation}

We will need the following easy lemma, which follows from dilatation
analyticity and Stone's theorem.

\begin{lemme}\label{lm:Dilatation}
Assume that the infrared cut-off parameter $\sigma$ is chosen such
that $g^2 \ll \sigma < g^{\frac{3}{2+\mu}} \ll 1.$ Then
\begin{equation}
\label{eq:Dilatation} \left ( \Psi_j , e^{-it H_g} f(H_g) \Psi_j
\right ) = A(t,\overline{\theta}) - A(t,\theta) +
B(t,\overline{\theta}) - B(t,\theta),
\end{equation}
for $\theta\in D(0,\theta_0), \ \ \Im\theta>0,$ where
\begin{align}
& A( t , \theta ) = \frac{1}{2  \pi i} \int_{\mathbb{R}} e^{-itz} f(z) \left ( \Psi_{j,\overline{\theta} } , \widetilde{R}_{g,\theta}^{ \sigma } (z) \Psi_{j,\theta} \right ) dz, \label{A(t,theta)} \\
&B( t , \theta ) = \frac{1}{2 \pi i} \int_{\mathbb{R}} e^{-itz} f(z)
\left ( \Psi_{j, \overline{\theta} } , \widetilde{R}_{g,\theta}^{
\sigma } (z) \sum_{n\geq1} \left ( - \widetilde{W}_{g,\theta}^{
\leqslant \sigma } \widetilde{R}_{g,\theta}^{ \sigma } (z) \right
)^n \Psi_{j,\theta} \right ) dz. \label{def_Btheta}
\end{align}
\end{lemme}

\begin{demo}.
By Stone's theorem,
\begin{equation}
\left ( \Psi_j , e^{-it H_g} f(H_g) \Psi_j \right ) = \lim_{
\varepsilon \searrow 0 } \frac{1}{2 \pi i} \int_{\mathbb{R}}
e^{-itz} f(z) \left ( \Psi_j , \left [ R_g( z - i\varepsilon ) -
R_g( z + i\varepsilon ) \right ] \Psi_j \right ) dz.
\end{equation}
Since $H_g$ and $\Psi_j$ are dilatation analytic, this implies for
$\theta\in D(0,\theta_0)$
\begin{equation}
\left ( \Psi_j , e^{-it H_g} f(H_g) \Psi_j \right ) = F( t ,
\overline{\theta} ) - F( t , \theta ),
\end{equation}
where
\begin{equation}
F( t , \theta ) = \frac{1}{2 \pi i} \int_{\mathbb{R}} e^{-itz} f(z)
\left ( \Psi_{j, \overline{\theta} } , R_{g,\theta} (z) \Psi_{j,
\theta}\right ) dz.
\end{equation}
 It follows from Lemma
\ref{lemma_Neumann}, below, that we can expand $R_{g,\theta}(z)$
into a Neumann series, which is convergent under our assumptions on
$g$ and $\sigma$ if Fermi's Golden Rule holds. We obtain
\begin{equation}
F( t , \theta ) = A( t , \theta ) + B( t , \theta ),
\end{equation}
for $\theta\in D(0,\theta_0), \ \ \Im\theta>0,$ and hence the claim
of the lemma is proven.
\end{demo}
\newline

In what follows, we fix $\theta\in D(0,\theta_0)$ with
$\Im\theta>0.$ We estimate $A(t,\overline{\theta})-A(t,\theta)$ and
$B( t , \overline{\theta} ) - B( t , \theta )$ in the following two
lemmata.

\begin{lemme}\label{lm:A_Estimate}
For $g^2\ll \sigma <\delta \ll 1,$ we have
\begin{equation*}
A(t,\overline{\theta}) - A(t,\theta)= e^{-it\lambda_{j,g}} +
O(\delta g^2\sigma^{-2}) + O(g^2\sigma^{-1}),
\end{equation*}
for all $t\ge 0.$
\end{lemme}
\begin{demo}.
It follows from the spectral theorem that
\begin{equation}\label{use_of_Lambda}
\mathcal{V} \widetilde{R}_{g,\theta}^\sigma (z) \mathcal{V}^{-1} =
\int_{\sigma(H_f^{\leqslant \sigma})}
\widetilde{R}_{g,\theta}^{\geqslant \sigma} (z-e^{-\theta}\omega)\otimes
dE_{H_f^{\leqslant \sigma}}(\omega),
\end{equation}
where $E_{H_f^{\leqslant \sigma}}$ are the spectral projections of
$H_f^{\leqslant \sigma}$; see for example \cite{RS4}. Furthermore,
$\mathcal{V} \Psi_{j,\theta} = \psi_{j,\theta} \otimes \Omega^{
\geqslant \sigma } \otimes \Omega^{ \leqslant \sigma }$, where
$\Omega^{ \geqslant \sigma }$ (respectively $\Omega^{ \leqslant
\sigma }$) denotes the vacuum in $\mathcal{F}_s^{ \geqslant \sigma
}$ (in $\mathcal{F}_s^{ \leqslant \sigma }$). Inserting this into
$(\ref{A(t,theta)})$ and using (\ref{use_of_Lambda}), we get
\begin{equation}
A( t , \theta ) = \frac{1}{2 i \pi} \int_{\mathbb{R}} e^{-itz} f(z)
\left ( \psi_{j, \overline{\theta} } \otimes \Omega^{ \geqslant
\sigma } , \widetilde{R}_{g,\theta}^{ \geqslant \sigma } (z)
\psi_{j,\theta } \otimes \Omega^{ \geqslant \sigma } \right ) dz.
\end{equation}
>From Proposition \ref{theorem_Hgeqsigma}, we know that the spectrum
of $H_{g,\theta}^{ \geqslant \sigma }$ is of the form pictured in
figure \ref{spectrum_Hgeqsigma}.
\begin{figure}[htbp]
\centering
\resizebox{0.8\textwidth}{!}{ \includegraphics{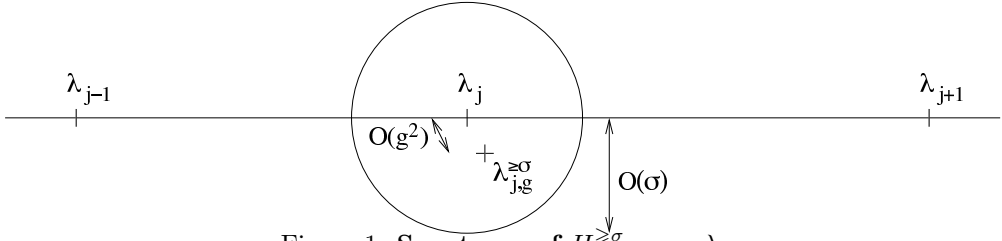} }
\caption{\textbf{Spectrum of $H_{g,\theta}^{ \geqslant \sigma }$
near $\lambda_{j}$ } } \label{spectrum_Hgeqsigma}
\end{figure}
In particular, a gap of order $\sigma$ opens between the
non-degenerate eigenvalue $\lambda_{j,g}^{ \geqslant \sigma }$ and
the essential spectrum of $H_{g,\theta}^{ \geqslant \sigma }$. By
Proposition \ref{lambda_diff_estimate}, the same holds for
$\widetilde{H}_{g,\theta}^{ \geqslant \sigma }$ instead of
$H_{g,\theta}^{ \geqslant \sigma }$, with $\lambda_{j,g}$ replacing
$\lambda_{j,g}^{ \geqslant \sigma }$, since $| \lambda_{j,g} -
\lambda_{j,g}^{ \geqslant \sigma } | \leq \mathrm{C}
g^2 \sigma^{1+\mu}$ and we assumed that $g^2 \ll \sigma \ll 1$.

Let us begin to estimate $A(t , \overline{\theta} ) - A(t , \theta
)$ by considering the contribution of the regular part,
$\widehat{R}_{g, \theta }^{ \geqslant \sigma } (z),$ in $A( t ,
\theta )$. By applying Green's theorem, we find that
\begin{equation}
\label{eq:RegularPart}
\begin{split}
R( t , \theta ) &:= \frac{1}{2 i \pi} \int_{\mathbb{R}} e^{-itz} f(z) \left ( \Psi_{j, \overline{\theta} } , \widehat{R}_{g, \theta }^{ \geqslant \sigma } (z) \Psi_{j, \theta }  \right ) dz \\
& = \frac{1}{2 i \pi} \int_{ \Gamma( \gamma_1 )} e^{-itz} \widetilde{f} (z) \left ( \Psi^{ \geqslant \sigma}_{j,\overline{\theta} } , \widehat{R}_{g, \theta }^{ \geqslant \sigma } (z) \Psi^{ \geqslant \sigma}_{j, \theta } \right ) dz \\
& \quad + \frac{1}{2 i \pi} \iint_{ \mathrm{D}( \gamma_1 ) }
e^{-itz} ( \partial_{ \bar{z} } \widetilde{f}) (z) \left ( \Psi^{
\geqslant \sigma}_{j, \overline{\theta} } , \widehat{R}_{g, \theta
}^{ \geqslant \sigma } (z) \Psi^{ \geqslant \sigma}_{j,\theta }
\right ) dzd\overline{z},
\end{split}
\end{equation}
where $\Psi^{ \geqslant \sigma}_{j, \theta } = \psi_{j, \theta }
\otimes \Omega^{ \geqslant \sigma },$ and $\Gamma( \gamma_1 )$ and
$\mathrm{D}( \gamma_1 )$ denote respectively the curve and the
domain pictured in figure \ref{curve_Gamma}, such that the interval
$\mathrm{I}_0$ strictly contains $\mathrm{I}$.
\begin{figure}[htbp]
\centering
\resizebox{0.8\textwidth}{!}{ \includegraphics{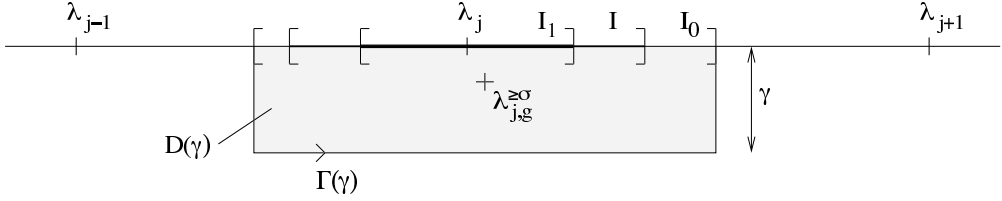} }
\caption{\textbf{Deformation of the path of integration} }
\label{curve_Gamma}
\end{figure}

By Proposition \ref{theorem_Hgeqsigma} and (\ref{eq:Regular}), the
regular part $\widehat{R}_{ g , \theta }^{ \geqslant \sigma } (z)$
in $(\ref{decomp_R_g,theta})$ is analytic in $z \in \mathrm{D}(
\gamma_1 )$ and satisfies
\begin{equation}\label{estimate_regular_part}
\left \| \widehat{R}_{ g , \theta }^{ \geqslant \sigma } (z) \right
\| \leq \frac{ \mathrm{C} }{ \mathrm{dist} ( z , \sigma (
\widetilde{H}_{g,\theta}^{ \geqslant \sigma } ) \setminus \{
\lambda_{j,g} \} ) },
\end{equation}
where $\mathrm{C}$ is a positive constant. We also have from
$(\ref{eq:SpectralDecomp})$  that
\begin{equation}
P_{0,\theta}^{\geqslant \sigma} \widehat{R}_{ g , \theta }^{
\geqslant \sigma } (z)P_{0,\theta}^{\geqslant \sigma} =
(P_{0,\theta}^{\geqslant \sigma}-P_{g,\theta}^{\geqslant
\sigma})\widehat{R}_{ g , \theta }^{ \geqslant \sigma }
(z)(P_{0,\theta}^{\geqslant \sigma}-P_{g,\theta}^{\geqslant
\sigma}),
\end{equation}
and from $(\ref{eq:SpectralProj})$ that
\begin{equation}
\label{eq:ProjState} P_{0,\theta}^{\geqslant \sigma} =
P_{0,\theta}^{\geqslant\sigma}P_{g,\theta}^{\geqslant \sigma}
P_{0,\theta}^{\geqslant \sigma} - (P_{g,\theta}^{\geqslant
\sigma}-P_{0,\theta}^{\geqslant \sigma})(P_{g,\theta}^{\geqslant
\sigma}-1)(P_{g,\theta}^{\geqslant \sigma}-P_{0,\theta}^{\geqslant
\sigma}),
\end{equation}
and from Proposition \ref{corollary_diff_projections}, below,
that
\begin{equation}
\label{eq:DiffProf} \|P_{g,\theta}^{\geqslant
\sigma}-P_{0,\theta}^{\geqslant \sigma}\| \le \mathrm{C}
g\sigma^{-1/2},
\end{equation}
for some positive constant C. Thus, by
$(\ref{eq:RegularPart})-(\ref{eq:DiffProf}),$ our assumptions on
$\widetilde{f}$ and the fact that $|I_0|=O(\delta),$ we get
\begin{equation}\label{regular_1}
\left | R( t , \theta ) \right |  = O(\delta g^2\sigma^{-2}e^{ -t
\gamma_1 } ) + O( |\gamma_1/\delta|^{n} ),
\end{equation}
where $0<\gamma_1 < \sigma \sin(\Im\theta) ,$ and any $n\in {\mathbb
N}.$ Similarly, for the contribution of the regular part
$\widehat{R}_{g,\overline{\theta}}^{ \geqslant \sigma }$ in $A( t ,
\overline{\theta} ),$ we have the following estimate
\begin{equation}\label{regular_2}
\left | R( t , \overline{\theta} ) \right | = O(\delta
g^2\sigma^{-2} e^{ -t \gamma_2 } ) + O( |\gamma_2/\delta|^{n} ),
\end{equation}
where $0<\gamma_2 < \sin(\Im\theta) \sigma,$ and any $n\in {\mathbb
N}.$

Next, we estimate the singular part of $A( t , \overline{\theta} ) -
A( t , \theta ).$ It is given by
\begin{equation}
S( t , \overline{\theta} ) - S( t , \theta ) := \frac{ C_g^\sigma (
\overline{\theta} ) }{2 i \pi} \int_{\mathbb{R}} e^{-itz} f(z) \left
( z - \overline{ \lambda_{j,g} } \right )^{-1} dz - \frac{
C_g^\sigma ( \theta ) }{ 2 i \pi} \int_{\mathbb{R}} e^{-itz} f(z)
\left ( z - \lambda_{j,g} \right )^{-1} dz,
\end{equation}
where we use the notation
\begin{equation}
C_g^\sigma ( \theta ) = \left ( \Psi_j^{ \geqslant \sigma } (
\overline{ \theta } ) , P_{g,\theta}^{ \geqslant \sigma } \Psi_{j,
\theta}^{ \geqslant \sigma }  \right ).
\end{equation}
By Proposition \ref{corollary_diff_projections}, we know that
\begin{equation}\label{Agsigma=1+O}
C_g^\sigma ( \theta ) = 1 + O(g^2 \sigma^{-1}).
\end{equation}
We deform the path of integration as we did above, adding a circle
$\mathcal{C}_\rho$ of radius $\rho$ around $\lambda_{j,g}$. This
yields
\begin{equation}\label{singular_part}
\begin{split}
S( t , \overline{\theta} ) - S( t , \theta ) = & \frac{1}{2 i \pi} \int_{ \Gamma (\gamma_3) }  e^{-itz} \widetilde{f} (z) \left [ \frac{ C_g^\sigma ( \overline{\theta} ) }{ z - \overline{ \lambda_{j,g} } } - \frac{ C_g^\sigma (\theta) }{ z - \lambda_{j,g} } \right ] dz \\
& + \frac{1}{2 i \pi} \int_{ \mathcal{C}_\rho }  e^{-itz} \widetilde{f} (z) \left [ \frac{ C_g^\sigma ( \overline{\theta} ) }{ z - \overline{ \lambda_{j,g} } } - \frac{ C_g^\sigma (\theta) }{ z - \lambda_{j,g} } \right ] dz \\
& + \frac{1}{2 i \pi} \iint_{ \mathrm{D} ( \gamma_3 ) \setminus
\mathrm{D}_\rho } e^{-itz} ( \partial_{ \bar{z} } \widetilde{f} )
(z) \left [ \frac{ C_g^\sigma ( \overline{\theta} ) }{ z -
\overline{ \lambda_{j,g} } } - \frac{ C_g^\sigma (\theta) }{ z -
\lambda_{j,g} } \right ] dz d\overline{z},
\end{split}
\end{equation}
for all $\rho>0$ sufficiently small, where $\mathrm{D}_\rho$ denotes
the disc of radius $\rho$ centered at $\lambda_{j,g},$ and
$0<\gamma_3 < \sin(\Im\theta)\sigma.$ The first integral can be
estimated by using arguments similar to those used to estimate the
regular part, (\ref{Agsigma=1+O}), and the fact that
\begin{equation*}
\Im\lambda_{j,g}=O(g^2).
\end{equation*}
We then obtain that for $0<g^2<\sigma\ll 1$
\begin{equation}\label{first_singular_integral}
\left | \frac{1}{2 i \pi} \int_{ \Gamma (\gamma_3) }  e^{-itz}
\widetilde{f} (z) \left [ \frac{ C_g^\sigma ( \overline{\theta} ) }{
z - \overline{ \lambda_{j,g} } } - \frac{ C_g^\sigma (\theta) }{ z -
\lambda_{j,g} } \right ] dz \right | = O( \delta g^2 \sigma^{-2}e^{
-t \gamma_3 } ).
\end{equation}
Similarly, since $( \partial_{ \overline{z} } \widetilde{f}) =0$ on
$\left \{ z | \ \  \mathrm{Re}(z) \in \mathrm{I}_1 \right \}$, we
see that the third integral in the r.h.s. of $(\ref{singular_part})$
is independent of $\rho$, for $\rho$ sufficiently small, and that
for any $n\in {\mathbb N}$
\begin{equation}
\left | \frac{1}{2 i \pi} \iint_{ \mathrm{D} ( \gamma_3 ) \setminus
\mathrm{D}_\rho } e^{-itz} ( \partial_{ \bar{z} } \widetilde{f} )
(z) \left [ \frac{ C_g^\sigma ( \overline{\theta} ) }{ z -
\overline{ \lambda_{j,g} } } - \frac{ C_g^\sigma (\theta) }{ z -
\lambda_{j,g} } \right ] dz \right | = O( |\gamma_3/\delta|^n ).
\end{equation}
It remains to estimate the second integral on the right hand side of
$(\ref{singular_part})$. Taking the limit as $\rho \rightarrow 0$
leads to the ``residue'' $C_g^\sigma ( \theta ) e^{ -it
\lambda_{j,g} } \widetilde{f}( \lambda_{j,g} )$. Since, by
construction, $\widetilde{f} = 1$ on $\left \{ z | \ \
\mathrm{Re}(z) \in \mathrm{I}_1 \right \}$, we get
\begin{equation}
\label{sing} \lim_{ \rho \rightarrow 0 } \frac{1}{2 i \pi} \int_{
\mathcal{C}_\rho }  e^{-itz} \widetilde{f} (z) \left [ \frac{
C_g^\sigma ( \overline{\theta} ) }{ z - \overline{ \lambda_{j,g} } }
- \frac{ C_g^\sigma (\theta) }{ z - \lambda_{j,g} } \right ] dz =
C_g^\sigma ( \theta ) e^{ -it \lambda_{j,g} }.
\end{equation}
The claim of the lemma follows from
$(\ref{regular_1})-(\ref{sing}).$
\end{demo}
\newline

\begin{lemme}\label{theorem_Btheta}
Assume that the infrared cut-off parameter $\sigma$ is chosen such
that $g^2 \ll \sigma \ll 1.$ Then, for all
times $t\ge 0,$ we have that
\begin{equation}
\left | B( t , \overline{\theta} ) - B( t , \theta ) \right | = O (
\delta g^{-2} \sigma^{\frac{1}{2}+\mu} ),
\end{equation}
where $B( t , \theta )$ is defined in $(\ref{def_Btheta}).$
\end{lemme}
\begin{demo}.
Recall that
\begin{equation}\label{Btheta_Neumann}
B( t , \theta ) = \sum_{n\geq1} B^n( t , \theta ),
\end{equation}
where
\begin{equation}
\label{eq:Bn} B^n( t , \theta )=\frac{1}{2 i \pi} \int_{\mathbb{R}}
e^{-itz} f(z) \left ( \Psi_j( \overline{\theta} ) ,
\widetilde{R}_{g,\theta}^{ \sigma } (z)  \left ( -
\widetilde{W}_{g,\theta}^{ \leqslant \sigma }
\widetilde{R}_{g,\theta}^{ \sigma } (z) \right )^n \Psi_j( \theta )
\right ) dz.
\end{equation}
It follows from \eqref{eq:Bn} and Lemma \ref{lemma_Neumann}
that\footnote{
Estimate $(\ref{eq:estimate_Bn})$ can be improved if one uses
instead of $\Psi_j$ a state that is a better approximation to the
resonance eigenstate.}
\begin{equation}\label{eq:estimate_Bn}
|B^n(t,\theta)| = O(\delta g^{-2} \sigma^{n( \frac{1}{2}+\mu ) } ),
\end{equation}
uniformly in $t\ge 0.$ Together with $(\ref{Btheta_Neumann})$ and
the assumption on $\sigma$ and $g,$ it follows that
\begin{equation}
|B(t,\theta)| = O(\delta g^{-2} \sigma^{\frac{1}{2}+\mu}),
\end{equation}
uniformly in $t.$ One can similarly show that
\begin{equation}
|B(t,\overline{\theta})| =O(\delta g^{-2} \sigma^{\frac{1}{2}+\mu} ),
\end{equation}
and hence the claim of the lemma follows. \end{demo}
\newline

\begin{demo} {\bf of Proposition \ref{pr:main}}.
It follows from Lemmata \ref{lm:Dilatation}, \ref{lm:A_Estimate} and
\ref{theorem_Btheta} that for $ t\ge 0$
\begin{equation}
\label{eq:MetaStab} \left ( \Psi_j , e^{-it H_g} f(H_g) \Psi_j
\right ) = e^{-it\lambda_{j,g}} + O(\delta g^2\sigma^{-2}) +
O(\delta g^{-2}\sigma^{1/2+\mu}) +
O(g^{2}\sigma^{-1}).
\end{equation}
Let $\delta = C\sigma,$ for some $C>1.$ We optimize the estimate on
the error term by choosing
\begin{equation}
\sigma = g^{2-\frac{2+4\mu}{5+2\mu}},
\end{equation}
and hence the claim of the proposition is proven.
\end{demo}
\newline

\begin{demo} {\bf  of Theorem \ref{main_theorem}}.
Proposition \ref{pr:main} implies that, for $t=0,$
\begin{equation}
(\Psi_j, (1-f(H_g)) \Psi_j) = \|\sqrt{1-f(H_g)}\Psi_j\|^2 =
O(g^{\frac{2+4\mu}{5+2\mu}}),
\end{equation}
which, together with the boundedness of the unitary operator
$e^{-itH_g}$ and Proposition \ref{pr:main}, for arbitrary $t>0,$
yields
\begin{align*}
(\Psi_j, e^{-itH_g} \Psi_j) &= (\Psi_j, e^{-itH_g} (1-f(H_g)+ f(H_g) )\Psi_j) \\
&= (\Psi_j, e^{-itH_g} f(H_g)\Psi_j) + O(\|\sqrt{1-f(H_g)}\Psi_j\|^2) \\
&= e^{-it\lambda_{j,g}} + O(g^{\frac{2+4\mu}{5+2\mu}}).
\end{align*}
\end{demo}

\section{The Hamiltonian $H_{g,\theta}^\sigma$} \label{sec:Ham}

In this section, we study the operator $H_{g,\theta}^\sigma$ used in
the previous section as an  approximation of $H_{g,\theta}$. We use
the Feshbach-Schur map\footnote{In  \cite{BFS1, BFS2, BCFS} this map
is called the Feshbach map. As was pointed out to us by F. Klopp and
B. Simon, the invertibility procedure at the heart of this map was
introduced by I. Schur in 1917; a similar approach was developed in
an independent work of H. Feshbach on the theory of nuclear
reactions in 1958, see \cite{GH} for further extensions and
historical remarks.}, \cite{BFS1,BFS2}, defined for a projection $P$
and a closed operator $H$ whose domain is contained in
$\mathrm{Ran}(P)$, by
\begin{equation}
\mathcal{F}_P(H) = PHP - P H \overline{P} \left [ \overline{P} H
\overline{P} \right ]^{-1} \overline{P} H P,
\end{equation}
where $\overline{P} = 1 - P$. Note that the domain of
$\mathcal{F}_P$ consists of operators $H$ such that
\begin{equation}
\left [ \overline{P} H \overline{P} \right ]^{-1}|_{\mathrm{Ran}
(\overline{P})}, \quad P H \overline{P} \left [ \overline{P} H
\overline{P} \right ]^{-1}|_{\mathrm{Ran} (\overline{P})}, \quad \left [
\overline{P} H \overline{P} \right ]^{-1} \overline{P} H P
\end{equation}
extend to bounded operators. We begin with the following
proposition.
\begin{prop}\label{theorem_Hgeqsigma}
Suppose $0<g^2\ll \sigma \ll 1.$ Then, for $\theta\in D(0,\theta_0)$
such that $\Im\theta\ne 0$ and $\sigma < \mathrm{d}_j \sin | \Im \theta |$, the spectrum of $H_{g,\theta}^{
\geqslant \sigma }$ in the disc $D( \lambda_{j} , \sigma/2 )$
consists of a single eigenvalue
\begin{equation}\label{loc_Hgeqsigma_1}
\sigma ( H_{g,\theta}^{ \geqslant \sigma } ) \cap
D(\lambda_{j},\sigma/2) = \{ \lambda_{j,g}^{ \geqslant \sigma } \}.
\end{equation}
Furthermore, there exists $\varepsilon>0$ such that, for all $z$ in
$D(\lambda_{j},\sigma/3)$ such that $\left | z - \lambda_{j,g}^{
\geqslant \sigma } \right | \gg g^{2+\varepsilon}$,
\begin{equation}\label{estimate_Rgeqsigma}
\left \| R_{g,\theta}^{ \geqslant \sigma }(z) \right \| \leq \frac{
\mathrm{C} }{ \mathrm{dist}( z , \sigma( H_{ g,\theta}^{ \geqslant
\sigma } ) ) },
\end{equation}
for some positive constant $\mathrm{C}$.
\end{prop}
\begin{demo}.
Let $P_\theta  := P_{0,\theta}^{ \geqslant \sigma } = P_{p, j, \theta } \otimes P_\Omega^{ \geqslant \sigma }$ and $\overline{P}_\theta := 1 - P_\theta$. For $\sigma < \mathrm{d}_j \sin | \Im \theta |$ and $z \in D(\lambda_{j} , \sigma/2 )$, one can show that, for any $n\geq 1$,
\begin{equation}\label{eq:estimate_Neumann}
\left \| \left [ \overline{P}_\theta  H_{0,\theta}^{ \geqslant \sigma } - z
\right ]^{-1} \left ( - W_{g,\theta}^{ \geqslant
\sigma } \left [ \overline{P}_\theta  H_{0,\theta}^{ \geqslant
\sigma } - z \right ]^{-1} \right )^n \right \| \leq \mathrm{C}_1 \sigma^{-1} \left (  \mathrm{C}_2 g \sigma^{-1/2} \right )^n,
\end{equation}
where $\mathrm{C}_1$, $\mathrm{C}_2$ are positive constant. Hence
for $g \sigma^{-1/2} \ll 1$ and any $z$ in $D( \lambda_{j} ,
\sigma/2 )$, the operator $\overline{P}_\theta H_{g,\theta}^{
\geqslant \sigma } \overline{P}_\theta  - z$ is invertible and its
inverse is given by the convergent Neumann series as
\begin{equation}\label{Neumann_Hgesigma}
\left [ \overline{P}_\theta H_{g,\theta}^{ \geqslant \sigma }
\overline{P}_\theta  - z \right ]^{-1}= \left [ \overline{P}_\theta
H_{0,\theta}^{ \geqslant \sigma } - z \right ]^{-1} \sum_{ n \geq 0
} \left ( - W_{g,\theta}^{ \geqslant \sigma } \left [
\overline{P}_\theta  H_{0,\theta}^{ \geqslant \sigma } - z \right
]^{-1} \right )^n .
\end{equation}
This implies that the operator $H_{g,\theta}^{ \geqslant \sigma } -
z $ is in the domain of $\mathcal{F}_{ P_\theta }$. Moreover,
\eqref{eq:estimate_Neumann} and \eqref{Neumann_Hgesigma} lead to the
estimates
\begin{equation}
\label{eq:ResEst} \left \| (H_f +1)^{n} \left [ \overline{P}_\theta
H_{g,\theta}^{ \geqslant \sigma } \overline{P}_\theta  - z \right
]^{-1} \right \| \le \mathrm{C}\sigma^{-1}, \ \ n=0,1,
\end{equation}
for some positive constant $\mathrm{C}$. 

Note that our choice of $P_\theta $ yields $P_\theta  W_{g,\theta}^{
\geqslant \sigma } P_\theta  = 0$. Therefore
\begin{equation}\label{eq_Feshbach}
\mathcal{F}_{ P_\theta  }( H_{g,\theta}^{ \geqslant \sigma } - z ) =
\left ( \lambda_{j} - z \right ) P_\theta  - P_\theta
W_{g,\theta}^{ \geqslant \sigma } \overline{P}_\theta  \left [
\overline{P}_\theta  H_{g,\theta}^{ \geqslant \sigma }
\overline{P}_\theta  - z \right ]^{-1} \overline{P}_\theta
W_{g,\theta}^{ \geqslant \sigma } P_\theta .
\end{equation}
The non-degeneracy of $\lambda_{j}$ implies that $\mathcal{F}_{
P_\theta  }( H_{g,\theta}^{ \geqslant \sigma } - z )$ can be written
as $[ \lambda_{j} - z + a(z)  ] P_\theta $, where $a(z)$ is a function from $D(\lambda_{j},\sigma/2) \rightarrow
\mathbb{C}$. Following \cite{BFS1,BFS3} (see also Proposition
\ref{theorem_diff_resonances} below), we have
\begin{equation}\label{eq:a(z)}
a(z) = g^2 Z_{j,\theta} + O( g^{2+\varepsilon} )
\end{equation}
for some $\varepsilon>0$, where $Z_{j,\theta}:= Z_{j,\theta}^{
\mathrm{od} } + Z_{j,\theta}^{ \mathrm{d} }$ with
\begin{align}
& Z_{j,\theta}^{ \mathrm{od} } =  \int_{ \mathbb{R}^3 } \mathcal{U}_\theta P_{p,j} \overline{G_x} (k) \overline{P}_{p,j} \left [ H_{p} - \lambda_{j} + \omega(k) - i0 \right ]^{-1} \overline{P}_{p,j} G_x (k) P_{p,j} \mathcal{U}_\theta^{-1} dk, \label{Z_j^od} \\
& Z_{j,\theta}^{ \mathrm{d} }  = \int_{ \mathbb{R}^3 }
\mathcal{U}_\theta P_{p,j} \overline{G_x} (k) P_{p,j} G_x (k)
P_{p,j} \mathcal{U}_\theta^{-1} \frac{ dk }{ \omega(k) }.
\label{Z_j^d}
\end{align}
Using the Leibniz rule and the fact that
\begin{equation}
\frac{d}{dz} \left [ \overline{P}_\theta  H_{0,\theta}^{ \geqslant
\sigma } - z \right ]^{-1} = \left [ \overline{P}_\theta
H_{0,\theta}^{ \geqslant \sigma } - z \right ]^{-2},
\end{equation}
one can prove, by differentiating \eqref{eq_Feshbach} with respect
to $z$, that $z \mapsto b(z) := \lambda_{j} - z + a(z) $ is an
analytic function on $D( \lambda_{j} , \sigma/2 )$, and that $|
db(z)/dz - 1 | < 1$, provided that $g^2 \sigma^{-1}$ is sufficiently
small. This implies that $b$ is a bijection on
$D(\lambda_{j},\sigma/2)$.

The isospectrality of the Feshbach map (see \cite{BFS1,BFS2}) tells
us that
$$z \in \sigma( H_{g,\theta}^{ \geqslant \sigma } ) \iff 0 \in \mathcal{F}_{ P_\theta  }( H_{g,\theta}^{ \geqslant \sigma } - z ) \iff b(z)=0.$$
On the other hand, it follows from the usual perturbation theory,
applied to the isolated non-degenerate eigenvalue $\lambda_{j}$ of
$H_{0,\theta}^{ \geqslant \sigma }$, that the spectrum of
$H_{g,\theta}^{ \geqslant \sigma }$ is not empty in $D( \lambda_{j}
, \sigma/2 )$, for $g$ sufficiently small. Hence there exists a
unique $\lambda_{j,g}^{ \geqslant \sigma }$ in $D( \lambda_{j} ,
\sigma/2)$ such that $b( \lambda_{j,g}^{ \geqslant \sigma } ) = 0$,
that is
\begin{equation}
\sigma ( H_{g,\theta}^{ \geqslant \sigma } ) \cap D( \lambda_{j} ,
\frac{ \sigma }{2} ) = \{ \lambda_{j,g}^{ \geqslant \sigma } \}.
\end{equation}

To prove \eqref{estimate_Rgeqsigma}, we use the following identity
(see \cite{BFS1}):
\begin{equation}
\begin{split}
\left [ H_{g,\theta}^{ \geqslant \sigma } - z \right ]^{-1} = & \left [ P_\theta  - \left [ \overline{ P }_\theta  \left ( H_{g,\theta}^{ \geqslant \sigma } - z \right ) \overline{P}_\theta \right ]^{-1} \overline{ P }_\theta
W_{g,\theta}^{ \geqslant \sigma } P_\theta  \right ] \left [ \mathcal{F}_{ P_\theta  }( H_{g,\theta}^{ \geqslant \sigma } - z ) \right ]^{-1}\\
& \times \left [ P_\theta  - P_\theta  W_{g,\theta}^{ \geqslant \sigma } \overline{P}_\theta  \left [ \overline{ P }_\theta  \left ( H_{g,\theta}^{ \geqslant \sigma } - z \right ) \overline{P}_\theta  \right ]^{-1} \right ] \\
& + \left [ \overline{ P }_\theta  \left ( H_{g,\theta}^{ \geqslant
\sigma } - z \right ) \overline{P}_\theta  \right ]^{-1} \overline{
P }_\theta ,
\end{split}
\end{equation}
which holds for $z$ in $\rho( H_{g,\theta}^{ \geqslant \sigma } )
\cap D( \lambda_{j} , \sigma/2 )$. The simple form of $\mathcal{F}_{
P_\theta }( H_{g,\theta}^{ \geqslant \sigma } - z ),$
(\ref{eq:ResEst}) and the fact that $| a(z) - a( \lambda_{j,g}^{
\geqslant \sigma } ) | = O(g^{2+\varepsilon})$ by \eqref{eq:a(z)}
lead to
\begin{equation}
\left \| \left [ H_{g,\theta}^{ \geqslant \sigma } - z \right ]^{-1}
\right \| \leq \mathrm{C}_1 \left ( \frac{ 1 + g^2 \sigma^{-1} }{ |
z - \lambda_{j,g}^{ \geqslant \sigma } | - \mathrm{C}_2
g^{2+\varepsilon} } + \sigma^{-1} \right ),
\end{equation}
for some positive constants $\mathrm{C}_1,\mathrm{C}_2$. Hence the
proposition is proven for $z$ in $D( \lambda_{j} , \sigma/3 )$ such
that $| z - \lambda_{j,g}^{ \geqslant \sigma } | \gg
g^{2+\varepsilon}$.
\end{demo}
\newline

Recall that, for $g \geq 0$, $P_{g,\theta}^{ \geqslant \sigma }$
denotes the projection onto the eigenspace associated with the
eigenvalue $\lambda_{j,g}^{ \geqslant \sigma }$ of $H_{g,\theta}^{
\geqslant \sigma }$.
\begin{prop}\label{corollary_diff_projections}
Let $g,\sigma$ as in Proposition \ref{theorem_Hgeqsigma} and choose
$\theta\in D(0,\theta_0)$ such that $\Im\theta\ne 0$.  Then, for $g$
small enough,
\begin{equation}
\left \| P_{g,\theta}^{ \geqslant \sigma } - P_{0,\theta}^{
\geqslant \sigma } \right \| \leq \mathrm{C}_0 g \sigma^{-1/2},
\end{equation}
where $\mathrm{C}_0$ is a positive constant.
\end{prop}
\begin{demo}.
Let $\mathcal{C}_j$ denote a circle centered at $\lambda_{j}$, with
radius $\sigma/3$, so that $\mathcal{C}_j \subset \rho (
H_{g,\theta}^{ \geqslant \sigma } )$. Since we have assumed $g^2 \ll
\sigma$, for $g$ sufficiently small, $\mathcal{C}_j$ contains both
$\lambda_{j,g}^{ \geqslant \sigma }$ and $\lambda_{j}$. Thus,
\begin{equation}\label{diff_projections_1}
P_{g,\theta}^{ \geqslant \sigma } - P_{0,\theta}^{ \geqslant \sigma
} = \frac{1}{2\pi i} \oint_{ \mathcal{C}_j } \left [ R_{g,\theta}^{ \geqslant \sigma
}(z) - R_{0,\theta}^{ \geqslant \sigma }(z) \right ] dz.
\end{equation}
We expand $R_{g,\theta}^{ \geqslant \sigma }(z)$ into a Neumann
series
\begin{equation}
R_{g,\theta}^{ \geqslant \sigma }(z) = R_{0,\theta}^{ \geqslant
\sigma }(z) \sum_{n\geq 0} \left ( - W_{g,\theta}^{ \geqslant \sigma
} R_{0,\theta}^{ \geqslant \sigma }(z) \right )^n.
\end{equation}
One can show by following the method of \cite{BFS3} that, for $n=0,1/2,1$,
\begin{equation}
\| ( H_f^{ \geqslant \sigma } )^n R_{0,\theta}^{ \geqslant \sigma } \| = O( \sigma^{-1+n} ).
\end{equation}
Hence, using that $a(G_{x,\theta}^{ \geqslant \sigma }) P_\Omega^{ \geqslant \sigma } = 0$ and $\| a(G_{x,\theta}^{ \leqslant \sigma }) (H_f^{\geqslant \sigma })^{-1/2} \overline{P}_\Omega^{ \geqslant \sigma } \| = O(1)$, 
we obtain that for
all $n \geq 1$,
\begin{equation}\label{estimate_R0geqsigma}
\left \| R_{0,\theta}^{ \geqslant \sigma }(z)  \left ( -
W_{g,\theta}^{ \geqslant \sigma } R_{0,\theta}^{ \geqslant \sigma
}(z) \right )^n \right \| \leq \frac{ \mathrm{C}_1 }{\sigma} \left (
\mathrm{C}_2 g \sigma^{-1/2} \right )^n,
\end{equation}
where $\mathrm{C}_1$ and $\mathrm{C}_2$ denote positive constants.
Inserting this in $(\ref{diff_projections_1})$ and using the fact
that the radius of $\mathcal{C}_j$ is equal to $\sigma/3$, we obtain
\begin{equation}
\left \| P_{g,\theta}^{ \geqslant \sigma } - P_{0,\theta}^{
\geqslant \sigma } \right \| = \frac{1}{2\pi}\left \| \oint_{ \mathcal{C}_j }
R_{0,\theta}^{ \geqslant \sigma }(z) \sum_{n\geq 1} \left ( -
W_{g,\theta}^{ \geqslant \sigma } R_{0,\theta}^{ \geqslant \sigma
}(z) \right )^n dz \right \| \leq \mathrm{C}_0 g \sigma^{-1/2},
\end{equation}
provided that $g \sigma^{-1/2}$ is sufficiently small. Hence the
proposition is proven.
\end{demo}
\newline

Using a renormalization group analysis, we will prove in Proposition
\ref{lambda_diff_estimate} below the following estimate of the
difference between the eigenvalues $\lambda_{j,g}$ and
$\lambda_{j,g}^{ \geqslant \sigma }$ of $H_{g,\theta}$ and
$H_{g,\theta}^\sigma$:
\begin{equation*}
\lambda_{j,g} - \lambda_{j,g}^{ \geqslant \sigma } = O( g^2 \sigma^{1+\mu} ),
\end{equation*}
for any $\sigma>0$. Here we prove a weaker estimate, which holds
only for $\sigma \gg g^2$, but which does not require the use of a
renormalization group analysis. Besides, it is sufficient to obtain
the statement of Theorem \ref{main_theorem}, with the slightly
weaker error term $O( g^{\frac{1+2\mu}{4+2\mu}} )$ for the Nelson
model, and $O(g^{-1/4})$ for the QED one.
\begin{prop}\label{theorem_diff_resonances}
Suppose $0<g^2\ll \sigma < g^{\frac{3}{2+\mu}}.$ Then
\begin{equation}
\lambda_{j,g} - \lambda_{j,g}^{ \geqslant \sigma } = O \left ( g^{ 4
- \frac{3}{2+\mu} } \right ).
\end{equation}
\end{prop}
\begin{demo}.
For $g$ and $\sigma$ small enough, we choose $\theta \in
D(0,\theta_0)$, $\mathrm{Im}( \theta ) \neq 0$, such that $0 < g^2 \ll \sigma
< \mathrm{d}_j \sin | \Im \theta | <
1.$ For $\rho$ such that $g^2 \ll \rho < \mathrm{d}_j \sin | \Im \theta |$, let $P_\theta := P_{p,j,\theta } \otimes
\mathbf{1}_{ H_f < \rho }$. Following \cite{BFS1,BFS3},
$\lambda_{j,g}$ satisfies
\begin{equation}\label{eq_lambdajg}
\begin{split}
\left | \lambda_{j,g} - \lambda_{j} - g^2 Z_{j,\theta} \right | \leq
\sum_{i=1}^5 \| \mathrm{Rem}_i \|,
\end{split}
\end{equation}
where $Z_{j,\theta}:= Z_{j,\theta}^{ \mathrm{od} } + Z_{j,\theta}^{
\mathrm{d} }$, with $Z_{j,\theta}^{ \mathrm{od} }$ and
$Z_{j,\theta}^{ \mathrm{d} }$ given by $( \ref{Z_j^od} )$-$(
\ref{Z_j^d} )$, and
\begin{align}
& \mathrm{Rem}_1 = P_\theta  W_{g,\theta} P_\theta , \\
& \mathrm{Rem}_2 = P_\theta  W_{g,\theta} \overline{P}_\theta  \left [ \overline{P}_\theta  H_{0,\theta} - \lambda_{j,g} \right ]^{-1} \overline{P}_\theta W_{g,\theta} P_\theta  - g^2 Q_\theta  \label{eq:Rem2} \\
& \mathrm{Rem}_3 = g^2 \left [ Q_\theta  - Z_{j,\theta} \right ],  \label{eq:Rem3} \\
& \mathrm{Rem}_4 = P_\theta  W_{g,\theta} \left ( \overline{P}_\theta  \left [ \overline{P}_\theta  H_{0,\theta} - \lambda_{j,g} \right ]^{-1} \overline{P}_\theta  W_{g,\theta} \right )^2 P_\theta , \\
& \mathrm{Rem}_5 = P_\theta  W_{g,\theta} \sum_{ n \geq 3 } \left (
\overline{P}_\theta  \left [ \overline{P}_\theta  H_{0,\theta} -
\lambda_{j,g} \right ]^{-1} \overline{P}_\theta  W_{g,\theta} \right
)^n P_\theta .
\end{align}
Here we have set
\begin{equation}\label{eq_Q}
Q_\theta = \int_{ \mathbb{R}^3 } P_\theta  \overline{ G_{x,\theta} }
(k) \left [ \frac{ \overline{P}_\theta ( \omega( k ) ) }{
H_{0,\theta} + e^{- \theta } \omega(k) - \lambda_{j,g} } \right ]
G_{x,\theta} (k) P_\theta  dk,
\end{equation}
where $\overline{P}_{\theta}( \omega( k )) := 1 - P_\theta
( \omega( k ))$, and $P_\theta ( \omega( k )) := P_{p,j,\theta }
\otimes \mathbf{1}_{ H_f + \omega(k) < \rho }$. Using the expression
\eqref{def_Gxtheta} of $G_{x,\theta}$ and estimates similar to
\cite[Lemmas IV.6-IV.12]{BFS1} or \cite[Lemma 3.16]{BFS3}, we claim
that
\begin{align}\label{bound1_Rem}
& \left \| \mathrm{Rem}_1 \right \| = O \left ( g \rho^{ 1+\mu }
\right ), \quad \left \| \mathrm{Rem}_2 \right \| = O \left ( g^2
\rho^{ 1 + \mu } \right ), \quad \left \| \mathrm{Rem}_3 \right \| =
O \left ( g^2 \rho \right ).
\end{align}
The first bound in \eqref{bound1_Rem} easily follows from
\begin{equation}\label{eq:estimate_aP}
\int_{ \mathbb{R}^3 } \| P_\theta \overline{G_{x,\theta}(k)} \otimes
a^*(k) \| dk = \int_{ \mathbb{R}^3 } \| G_{x,\theta}(k) \otimes a(k)
P_\theta  \| dk \leq \mathrm{C} \rho^{1+\mu}.
\end{equation}
The second one follows from normal-ordering \eqref{eq:Rem2} and
using again \eqref{eq:estimate_aP}. Finally, the last bound in
\eqref{bound1_Rem} follows from computing the difference in
\eqref{eq:Rem3} and using the estimate
\begin{equation}\label{eq:barP/H0+omega}
\left \| \frac{ \overline{P}_\theta ( \omega( k )) }{ H_{0,\theta} +
e^{-\theta} \omega(k) - \lambda_{j,g} } \right \| \leq \frac{
\mathrm{C}_1 }{ ( \mathrm{d}_j \sin |\mathrm{Im} \theta| ) \omega(k)
},
\end{equation}
for some positive constant $\mathrm{C}_1$. Now it is proved in
\cite{BFS1,BFS3} that $\left \| \mathrm{Rem}_4 + \mathrm{Rem}_5
\right \| = O \left ( g^3 \rho^{-1/2} \right )$. Let us estimate
these terms more precisely: we claim that
\begin{align}\label{bound2_Rem}
& \left \| \mathrm{Rem}_4 \right \| = O \left ( g^3 \rho^{  \mu }
\right ), \quad \left \| \mathrm{Rem}_5 \right \| = O \left ( g^4
\rho^{ -1 } \right ).
\end{align}
To prove the first bound in \eqref{bound2_Rem}, we decompose
$W_{g,\theta}$ into $W_{g,\theta} = g ( a^*( G_{x,\theta} ) + a(
G_{x,\theta} ) )$ and estimate each term separately by normal
ordering. For instance, let us compute
\begin{equation}\label{eq:estimate_rem4}
\begin{split}
& P_\theta  a( G_{x,\theta} ) \frac{ \overline{P}_\theta  }{ H_{0,\theta} - \lambda_{j,g} } a^*( G_{x,\theta} ) \frac{ \overline{P}_\theta  }{ H_{0,\theta} - \lambda_{j,g} } a( G_{x,\theta} ) P_\theta  \\
& = P_\theta  \int_{ \mathbb{R}^9 } G_{x,\theta}( k_1 ) \otimes a( k_1 ) \frac{ \overline{P}_\theta  }{ H_{0,\theta} - \lambda_{j,g} } \overline{ G }_{x,\theta} ( k_2 ) \otimes a^*( k_2 ) \frac{ \overline{P}_\theta  }{ H_{0,\theta} - \lambda_{j,g} } \\
& \phantom{~ = P_\theta  \int_{ \mathbb{R}^9 } } G_{x,\theta} ( k_3
) \otimes a( k_3 ) P_\theta  dk_1 dk_2 dk_3.
\end{split}
\end{equation}
It follows from a pull-through formula and the canonical commutation
rules that the ``worst'' term we have to estimate from the rhs of
\eqref{eq:estimate_rem4} is
\begin{equation}\label{eq:Trho_theta}
\begin{split}
\mathrm{T}( \rho , \theta ) := P_\theta  \int_{ \mathbb{R}^6 } & G_{x,\theta}( k_1 ) \otimes
I \frac{ \overline{P}_\theta ( \omega( k_1 )) }{ H_{0,\theta} + e^{-\theta} \omega(k_1) - \lambda_{j,g} }
\overline{ G }_{x,\theta}( k_1 )  \\
& \otimes I \frac{ \overline{P}_\theta  }{ H_{0,\theta} -
\lambda_{j,g} } G_{x,\theta} ( k_3 ) \otimes a( k_3 ) P_\theta  dk_1
dk_3.
\end{split}
\end{equation}
One can see that
\begin{align}
& \left \| \frac{ \overline{P}_\theta  }{ H_{0,\theta} -
\lambda_{j,g} } \right \| \leq \frac{ \mathrm{C}_1 }{ ( \mathrm{d}_j
\sin |\mathrm{Im} \theta| ) \rho },
\end{align}
for some positive constant $\mathrm{C}_1$. Inserting this together
with \eqref{eq:barP/H0+omega} into \eqref{eq:Trho_theta}, we get
\begin{equation}
\begin{split}
\left \| \mathrm{T}( \rho , \theta ) \right \| & \leq \frac{ \mathrm{C}_1^2 }{ ( \mathrm{d}_j \sin |\mathrm{Im} \theta| )^2 } \rho^{-1} \int_{ \mathbb{R}^3 } \frac{ | G_{x,\theta}(k_1) |^2 }{ \omega( k_1 ) } dk_1\int_{ \mathbb{R}^3 } \| G_{x,\theta}(k_3) \otimes a(k_3) P_\theta  \| dk_3 \\
& \leq \frac{ \mathrm{C}_2 }{ ( \mathrm{d}_j \sin |\mathrm{Im} \theta|
)^2 } \rho^{\mu},
\end{split}
\end{equation}
where $\mathrm{C}_2$ is a positive constant. Since the other terms
could be estimated in the same way, the first bound in
\eqref{bound2_Rem} follows; the second bound in \eqref{bound2_Rem}
can be obtained by using similar computations (see also \cite[Lemma
3.16]{BFS3}).

For $\rho>\sigma$ the eigenvalue $\lambda_{j,g}^{ \geqslant \sigma
}$ of $H_{g,\theta}^{ \geqslant \sigma }$ is given by the formulas
\eqref{eq_lambdajg}$-$\eqref{eq_Q}, except that $W_{g,\theta}$ and
$G_{x,\theta}(k)$ are replaced respectively by $W_{g,\theta}^{
\geqslant \sigma }$ and $G_{x,\theta}^{ \geqslant \sigma }(k)$. For the
terms analogous to  $Z_{j,\theta}^{ \mathrm{od} }$ and
$Z_{j,\theta}^{ \mathrm{d} }$ we have by a straightforward
computations that
\begin{align}
& \int_{ |k| \leq \sigma } \mathcal{U}_\theta P_{p,j} \overline{G_x} (k) \overline{P}_{p,j} \left [ H_{p} - \lambda_{j} + \omega(k) - i0 \right ]^{-1} \overline{P}_{p,j} G_x (k) P_{p,j} \mathcal{U}_\theta^{-1} dk = 0, \label{diff_Zod} \\
& \int_{ |k| \leq \sigma } \mathcal{U}_\theta P_{p,j} \overline{G_x}
(k) P_{p,j} G_x (k) P_{p,j} \mathcal{U}_\theta^{-1} \frac{ dk }{
\omega(k) } = O \left ( \sigma^{1+2\mu} \right ), \label{diff_Zd}
\end{align}
where in \eqref{diff_Zod} we used the fact that
$\sigma<\mathrm{d}_j$. Hence, with the obvious notation,
$Z_{j,\theta}= Z_{j,\theta}^{ \geqslant \sigma }
+O(\sigma^{1+2\mu}).$ Furthermore, Eqns.
\eqref{bound1_Rem}$-$\eqref{bound2_Rem} still hold for
$\lambda_{j,g}^{ \geqslant \sigma }$.
Hence remembering the assumptions $\sigma < \rho$, $g^2 \ll \rho$ we
obtain
\begin{equation}\label{first_result_difference_resonances}
\lambda_{j,g} - \lambda_{j,g}^{ \geqslant \sigma } = O \left ( g
\rho^{1+\mu} \right ) + O \left ( g^2 \rho \right ) + O \left ( g^4
\rho^{-1} \right ).
\end{equation}
Optimizing with respect to $\rho$ leads to the claim of the
proposition.
\end{demo}
\newline

The following lemma was used in the proof of  Lemmata
\ref{lm:Dilatation} and \ref{theorem_Btheta}.
\begin{lemme}\label{lemma_Neumann}
Let $\theta$ in $D(0,\theta_0)$, $\Im \theta > 0$ and let $g,\sigma$
be such that $0<g^2 \ll \sigma \ll 1$. Then for
all $z \in \mathrm{I}$ and $n \geq 1$, we have the estimate:
\begin{equation}
\left \| \widetilde{R}_{g,\theta}^{ \sigma }(z) \left (
\widetilde{W}_{g,\theta}^{ \leqslant \sigma }
\widetilde{R}_{g,\theta}^{ \sigma }(z) \right )^n \right \| \leq
\mathrm{C}_1g^{-2} \left ( \mathrm{C}_2 \sigma^{1/2 + \mu } \right )^n,
\end{equation}
where $\mathrm{C}_1,\mathrm{C}_2$ are positive constants.
\end{lemme}
\begin{demo}.
Recall that
\begin{equation}\label{eq:lemma_Wleq}
\widetilde{W}_{g,\theta}^{ \leqslant \sigma } = g a^*(
G_{x,\theta}^{ \leqslant \sigma } ) + g a( G_{x,\theta}^{ \leqslant
\sigma } ) + \left ( \lambda_{j,g} - \lambda_{j,g}^{ \geqslant
\sigma } \right ).
\end{equation}
From the spectral representation
\eqref{use_of_Lambda} and the decomposition
$(\ref{decomp_R_g,theta})$, we can write
\begin{equation} \label{eq:RtildeDecomp}
\widetilde{R}_{g,\theta}^{ \sigma }(z) =
\widehat{R}_{g,\theta}^\sigma (z) + (
P_{g,\theta}^{ \geqslant \sigma } \otimes I ) \left [ z
- \lambda_{j,g}^{ \geqslant \sigma } - e^{-\theta} H_f^{ \leqslant \sigma }
\right ]^{-1},
\end{equation}
where $\widehat{R}_{g,\theta}^\sigma (z):= (I- ( P_{g,\theta}^{
\geqslant \sigma } \otimes I ) ) \widetilde{R}_{g,\theta}^{ \sigma
}(z)$. With the definition \eqref{3.7}, we have
\begin{equation}\label{eq:rep_rhat}
( H_f^{ \leqslant \sigma } )^n  \widehat{R}_{g,\theta}^\sigma (z) =
\int_{\sigma(H_f^{\leqslant \sigma})} \omega^n
\widehat{R}_{g,\theta}^{\geqslant \sigma}
(z-e^{-\theta}\omega)\otimes dE_{H_f^{\leqslant \sigma}}(\omega).
\end{equation}
It follows from Proposition \ref{theorem_Hgeqsigma} that for all $z$
in $\mathrm{I}$ and for $n=0, 1/2, 1$
\begin{equation}\label{RhatEst} \left \|
( H_f^{ \leqslant \sigma }  )^n \widehat{R}_{g,\theta}^\sigma
(z) \right \| = O( \sigma^{-1+n} ).
\end{equation}
Besides, for $n=0, 1/2, 1$
\begin{equation}\label{RsingEst}
\left \| ( H_f^{ \leqslant \sigma } )^n \left [ z -
\lambda_{j,g}^{ \geqslant \sigma } - e^{-\theta} H_f^{ \leqslant \sigma }
\right ]^{-1} \right \| \leq \left | \mathrm{Im} ( \lambda_{j,g}^{
\geqslant \sigma } ) \right |^{-1+n} = O(g^{-2(1-n)}),
\end{equation}
provided that Fermi's Golden Rule holds. 
From $a( G_{x,\theta}^{ \leqslant \sigma } ) P_\Omega^{ \leqslant \sigma } = 0$ and $ \| a( G_{x,\theta}^{ \leqslant \sigma } ) (H_f^{ \leqslant \sigma })^{-1/2} \overline{P}_\Omega^{ \leqslant \sigma } \| = O( \sigma^{1/2+\mu})$ (where $P_\Omega^{ \leqslant \sigma }$ denotes the projection onto the vacuum in $\mathcal{F}_s^{ \leqslant \sigma }$), we get
\begin{equation}\label{eq:lemma_a_a*}
\left \| a( G_{x,\theta}^{ \leqslant \sigma } ) \left [ z - \lambda_{j,g}^{ \geqslant \sigma } - e^{-\theta} H_f^{ \leqslant \sigma }  \right ]^{-1} \right \| 
= O( g^{-1} \sigma^{1/2+\mu} ).
\end{equation}
Similarly \eqref{RhatEst} leads to
\begin{equation}\label{eq:lemma_a_a*_2}
\left \| a( G_{x,\theta}^{ \leqslant \sigma } )
\widehat{R}_{g,\theta}^\sigma (z) \right \| = O( \sigma^\mu ).
\end{equation}
The claim of the lemma then follows from
$(\ref{eq:lemma_Wleq})-(\ref{eq:lemma_a_a*_2})$, the assumption $g^2
\ll \sigma <g^{\frac{3}{2+\mu}}$, and Proposition
\ref{lambda_diff_estimate}.
\end{demo}

\section{Extension to non-relativistic QED}\label{sec:QEDproof}

Now we extend the analysis above to the standard Hamiltonian of
non-relativistic QED introduced in (\ref{eq:NRQED}), Section
\ref{sec:Introduction}. Let now $H_{g,\theta}$ be the dilatation
deformation of the Hamiltonian $H^{SM}_g$ defined in \eqref{I.10}.
We keep the notation of Sections \ref{sec:Hamiltonians} -
\ref{sec:Ham}.

The results and proofs of Sections \ref{sec:MainResult} -
\ref{sec:Ham} go through without a change
except for the proof of Lemma \ref{theorem_Btheta}. 
 In the non-relativistic QED case, $W_{g,\theta}$ is given by
\begin{equation} \label{Wqed1}
W_{g,\theta} = g e^{-\theta} p \cdot A_\theta( x ) + \frac{g^2}{2}
A_\theta(x) \cdot A_\theta(x) - \frac{g^2}{2} \Lambda,
\end{equation}
where we used the notation $p \cdot A_\theta( x ) := -i \sum_{j=1}^N
\frac{1}{m_j}\nabla_j \cdot \mathcal{U}_\theta A(x_j)
\mathcal{U}_\theta^{-1}$, and similarly for $A_\theta(x) \cdot
A_\theta(x)$. The quantized vector potential $A(x_j)$ is given by
(\ref{eq:VectorPot}), and the constant $\Lambda$ is given by
$\Lambda := \frac{1}{ (2\pi)^3 } (\sum_j \frac{1}{m_j }) \int
\chi(k)^2/|k| d^3k $. Here we have
\begin{equation}\label{Wqed2}
W^{\leqslant\sigma}_{g,\theta} = ge^{-\theta} p \cdot A^{\leqslant
\sigma}_\theta (x) + g^2 A^{\leqslant \sigma}_\theta(x)
\cdot A^{\geqslant \sigma}_\theta(x) + \frac{g^2}{2} A^{ \leqslant
\sigma }_\theta(x) \cdot A^{ \leqslant \sigma}_\theta(x) -
\frac{g^2}{2} \Lambda^{ \leqslant \sigma },
\end{equation}
where $\Lambda^{ \leqslant \sigma } := \frac{1}{(2\pi)^3} (\sum_j \frac{1}{m_j}) \int_{ |k| \leq \sigma }
\chi(k)^2 / |k| d^3k$. Hence the QED Hamiltonian satisfies the
condition similar to \eqref{estimate_Wgleq} with $\mu = 0$. We show
now how to overcome this difficulty (a different way to proceed is
to use the Pauli-Fierz transform \cite{BFS1,BFS3,Sigal}).

In our sketch of the proof of Lemma \ref{theorem_Btheta}, we begin
with the most singular term, $B^1$, of the expansion
\eqref{Btheta_Neumann} in Section \ref{sec:MainResult}. Thus we have
to bound the term  $\widetilde{R}^\sigma_{g,\theta}
\widetilde{W}^{\leqslant \sigma}_{g,\theta}
\widetilde{R}^\sigma_{g,\theta}$. The part of
$\widetilde{W}^{\leqslant \sigma}_{g,\theta}$  involving the
difference of the eigenvalues is estimated in the same way as
before. Namely, using that $\| \widetilde{R}_{g,\theta}^\sigma \| =
O(g^{-2})$ and that, by Proposition \ref{lambda_diff_estimate}, $|
\lambda_{j,g} - \lambda_{j,g}^{ \geqslant \sigma } | = O( g^2 \sigma
)$, we obtain that
\begin{equation} \label{EVdifferContr}
\left \| \widetilde{R}^\sigma_{g,\theta}
 \left ( \lambda_{j,g} - \lambda_{j,g}^{ \geqslant \sigma } \right )
\widetilde{R}^\sigma_{g,\theta} \right \| = O( \sigma g^{-2} ).
\end{equation}

Now we estimate the remaining part $\widetilde{R}^\sigma_{g,\theta}
W^{\leqslant \sigma}_{g,\theta} \widetilde{R}^\sigma_{g,\theta}$ of
$\widetilde{R}^\sigma_{g,\theta} \widetilde{W}^{\leqslant
\sigma}_{g,\theta} \widetilde{R}^\sigma_{g,\theta}$. Using the
relation $ e^{-\theta}p_j =m_je^{\theta} i[H^\sigma_{g,\theta}
,x_j]-g A_\theta(x_j)$ the term $W_{g,\theta}^{ \leqslant \sigma }$
can be written as
\begin{align}\label{Wqed2}
W^{\leqslant\sigma}_{g,\theta} &= g e^{\theta}i[H^\sigma_{g,\theta}
,x] \cdot A^{\leqslant \sigma}_\theta (x) + I,
\end{align}
where $[H^\sigma_{g,\theta} ,x] \cdot A^{\leqslant \sigma}_\theta
(x) := \sum_{j} [H^\sigma_{g,\theta} ,x_{j}] \cdot A^{\leqslant
\sigma}_\theta (x_{j})$ and $I:= \frac{g^2}{2} \left [ A^{\leqslant
\sigma}_\theta (x) \cdot  A^{\leqslant \sigma}_\theta (x) - \Lambda^{
\leqslant \sigma } \right ]$. Furthermore, using that
$[H^\sigma_{g,\theta} ,x] \cdot A^{\leqslant \sigma}_\theta
(x)=[H^\sigma_{g,\theta} ,x \cdot A^{\leqslant \sigma}_\theta (x)]-
x \cdot [H^\sigma_{g,\theta} ,A^{\leqslant \sigma}_\theta (x)]$, we obtain
\begin{equation}\label{Wqed4}
W^{\leqslant\sigma}_{g,\theta} = g [H^\sigma_{g,\theta}
,x \cdot A^{\leqslant \sigma}_\theta (x)] + I + II,
\end{equation}
where $II:=- g x \cdot [H^\sigma_{g,\theta} ,A^{\leqslant \sigma}_\theta
(x)]$.  We can now rewrite the operator
$\widetilde{R}^\sigma_{g,\theta} W^{\leqslant \sigma}_{g,\theta}
\widetilde{R}^\sigma_{g,\theta} $ as
\begin{align} \label{Wqed5}
g x\cdot A^{\leqslant
\sigma}_\theta(x)\widetilde{R}^\sigma_{g,\theta} - g
\widetilde{R}^\sigma_{g,\theta}   x\cdot  A^{\leqslant
\sigma}_\theta (x)
 +\widetilde{R}^\sigma_{g,\theta} (I+ II )\widetilde{R}^\sigma_{g,\theta}.
\end{align}

Let $f$ be a (vector-)function of $k$. To estimate the expression
above we will use the following estimates
\begin{align} \label{StandardEst1}
& \left \| a( fG_{x,\theta}^{ \leqslant \sigma } )^n \psi \right \|
\leq
\mathrm{C} \sigma^{ n/2 }\sup_{|k| \le \sigma}|f| \left \| ( H_f^{ \leqslant \sigma } )^{n/2} \psi \right \|,  \quad n=1,2, \\
& \left \| a^*( G_{x,\theta}^{ \leqslant \sigma } ) \psi \right \| \leq
\mathrm{C} \left(\sigma^{ 1/2} \left \| ( H_f^{ \leqslant \sigma } )^{1/2} \psi \right \| + \sigma^{ } \| \psi  \|\right)
, \label{StandardEst2} \\
& \left \| ( H_f^{ \leqslant \sigma } )^n
\widetilde{R}_{g,\theta}^\sigma \right \| \leq \mathrm{C}
g^{2(n-1)}, \quad n=0,1/2,1.  \label{RtildeEst}
\end{align}
The first two estimates are standard (see e.g. \cite{BFS1,BFS3}). To
prove the last inequality one uses Eqns \eqref{eq:RtildeDecomp} -
\eqref{RsingEst}. In addition we need the following estimate for any
$\delta \ll | \lambda_j - \Sigma |$
\begin{equation}\label{eq:expdelta_tildeR}
\left \| ( H_f^{ \leqslant \sigma } )^n e^{ - \delta \langle x \rangle }
\widetilde{R}_{g,\theta}^\sigma e^{ \delta \langle x \rangle } \right \| \leq \mathrm{C} g^{2(n-1)}, \quad n=0,1/2,1,
\end{equation}
where, recall, $\Sigma = \inf \sigma_{\mathrm{ess}} (H_p)$, and
$\langle x \rangle := \sum_j [1+x_j^2]^{1/2}$. Eqn
\eqref{eq:expdelta_tildeR} follows in the same way as
\eqref{RtildeEst}, provided we prove that \eqref{estimate_Rgeqsigma}
still holds if one replaces $H_{g,\theta}^{ \geqslant \sigma }$ and
$R_{g,\theta}^{ \geqslant \sigma }$ respectively by $e^{- \delta
\langle x \rangle } H_{g,\theta}^{ \geqslant \sigma } e^{  \delta
\langle x \rangle }$ and $e^{ - \delta \langle x \rangle }
R_{g,\theta}^{ \geqslant \sigma } e^{ \delta \langle x \rangle }$.
To prove the latter property, we note that
\begin{align}
& W_{g,\theta}^{ \delta , \geqslant \sigma } = W_{g,\theta}^{ \geqslant \sigma } + i g \delta e^{-\theta} \sum_j \frac{1}{m_j} \langle x_j \rangle^{-1} x_j \cdot A^{ \geqslant \sigma }_\theta(x_j), \\
& H_{0,\theta}^{ \delta , \geqslant \sigma } = e^{- \delta \langle x
\rangle } H_{p,\theta} e^{ \delta \langle x \rangle } + e^{-\theta}
H_f^{ \geqslant \sigma },
\end{align}
where $ W_{g,\theta}^{ \delta , \geqslant \sigma }:= e^{ - \delta
\langle x \rangle } W_{g,\theta}^{  \geqslant \sigma } e^{ \delta
\langle x \rangle }$ and similarly for $H_{0,\theta}^{ \delta ,
\geqslant \sigma }$, $\overline{P}^\delta_\theta$, $P_{p, j, \theta
}^\delta$ and $ \overline{P}_{p, j, \theta }^\delta $. Using that
$\overline{P}^\delta_\theta = P_{p, j, \theta }^\delta \otimes
\overline{P}_\Omega^{ \geqslant \sigma }+ \overline{P}_{p, j, \theta
}^\delta \otimes I $ and the fact that $e^{- \delta \langle x
\rangle } H_{p,\theta} e^{  \delta \langle x \rangle }$ has the same
eigenvalues as $ H_{p,\theta}$ we write
$$\left  [   H_{0,\theta}^{ \delta , \geqslant \sigma } - z \right
]^{-1}\overline{P}^\delta_\theta= \left [ e^{-\theta} H_f^{
\geqslant \sigma }+ \lambda_j - z \right ]^{-1} (P_{p, j, \theta
}^\delta \otimes \overline{P}_\Omega^{ \geqslant \sigma })$$
\begin{equation}\label{eq:HoDecomp}
+ \left [
 H_{0,\theta}^{ \delta , \geqslant \sigma
} - z \right ]^{-1}(\overline{P}_{p, j, \theta }^\delta \otimes I).
\end{equation}
Using this decomposition we conclude, similarly to
\eqref{eq:estimate_Neumann}, that for $\sigma < \mathrm{d}_j \sin |
\Im \theta |$ and $z$ in $D(\lambda_j , \sigma/2)$, we have the for
some positive constants $\mathrm{C}_1$, $\mathrm{C}_2$
\begin{equation}\label{eq:estimate_Neumann_QED}
\left \| \left [ \overline{P}^\delta_\theta  H_{0,\theta}^{ \delta ,
\geqslant \sigma } - z \right ]^{-1} \left ( - W_{g,\theta}^{ \delta
, \geqslant \sigma } \left [ \overline{P}^\delta_\theta
H_{0,\theta}^{ \delta , \geqslant \sigma } - z \right ]^{-1} \right
)^n \right \| \leq \mathrm{C}_1 \sigma^{-1} \left (  \mathrm{C}_2 g
\sigma^{-1/2} \right )^n.
\end{equation}

Now, the first two terms in Eqn \eqref{Wqed5} have only one
resolvent each. Using estimates Eqns \eqref{StandardEst1} and
\eqref{StandardEst2}, with $n=1$ and $f\equiv 1$, and Eqns
\eqref{RtildeEst}-\eqref{eq:expdelta_tildeR} with $n=1/2$, we obtain
for these terms, times $e^{-\delta \langle x\rangle}$, with
$\delta>0$, the estimate $O( \sigma^{\frac{1}{2}}+ g^{-1}
\sigma^{})$. The operator
\begin{equation}
\label{eq:II} II= ige^{-\theta}x \cdot (e^{-\theta}p-xg A^{\leqslant
\sigma}_\theta (x))\nabla A^{\leqslant \sigma}_\theta (x)-g
e^{-2\theta}x\sum \frac{1}{2m_j}\Delta_j A^{\leqslant \sigma}_\theta
(x)+x[H^{\leqslant \sigma}_{f} ,A^{\leqslant \sigma}_\theta (x)]
\end{equation}
has better infrared behaviour than the original operator
$A^{\leqslant \sigma}_\theta (x)$ by an extra factor $k,\ \omega^2$
or $\omega$, which, due to \eqref{StandardEst1}, with $n=1$ and $f =
\omega\ \mbox{or}\ k$, and
\eqref{RtildeEst}-\eqref{eq:expdelta_tildeR}, gives the estimate $
\|e^{-\delta \langle x\rangle} \widetilde{R}^\sigma_{g,\theta}II
\widetilde{R}^\sigma_{g,\theta}\| = O(g^{-2} \sigma^{\frac{3}{2}})$.
Finally, the term $I$ is quadratic in $A^{\leqslant \sigma}_\theta
(x)$. Putting it to the normal form and using the estimates
\eqref{StandardEst1} and \eqref{RtildeEst} leads to the estimate
$\widetilde{R}^\sigma_{g,\theta} I \widetilde{R}^\sigma_{g,\theta} =
O (\sigma^{})$. Collecting the above estimates and using that $O(
\sigma^{\frac{1}{2}}+ g^{-1} \sigma^{}+ g^{-2} \sigma^{3/2}) = O(
g^{-2} \sigma^{3/2})$, we arrive at
\begin{equation}
\label{eq:EstB1Factor} \left \| e^{-\delta \langle x\rangle}
\widetilde{R}^\sigma_{g,\theta} W^{\leqslant \sigma}_{g,\theta}
\widetilde{R}^\sigma_{g,\theta} \right \|   = O( g^{-2}
\sigma^{3/2}).
\end{equation}
Next,  we pull $e^{-\delta \langle x\rangle}$, with $\delta>0$
sufficiently small, from $\Psi_j$
and use the above estimate to obtain
\begin{equation*}
\left | (\Psi_{j,\overline{\theta}},\widetilde{R}^\sigma_{g,\theta} W^{\leqslant \sigma}_{g,\theta}
\widetilde{R}^\sigma_{g,\theta}\Psi_{j,\theta}) \right |  = O(g^{-2}\sigma^{3/2}).
\end{equation*}
Therefore, the largest contribution to $B^1$ comes from the term
\eqref{EVdifferContr} that involves the difference of the
eigenvalues. Taking into account the factor $\sigma$ obtained from
the $z$ integration yields that $B^1= O(\sigma^2 g^{-2}).$

One can estimate the operators $B^n,\ n\ge 2,$ similarly. In
particular, we claim that $B^n = O(\sigma^{ \frac{n + 4}{2}}
g^{-2})$ for $n \geq 2$. Consider for example the term $B^2.$ Since
$\widetilde{R}^\sigma_{g,\theta}= O(g^{-2})$ and
$(\lambda_{j,g}-\lambda_{j,g}^{\geqslant\sigma})
P_{g,\theta}^{\geqslant \sigma}\otimes I = O(g^2 \sigma)$, we have
that
\begin{equation}
\label{eq:B2Est1} \big
\|\widetilde{R}^\sigma_{g,\theta}(\lambda_{j,g}-\lambda_{j,g}^{\geqslant\sigma})
(P_{g,\theta}^{\geqslant \sigma}\otimes I)
\widetilde{R}^\sigma_{g,\theta}
(\lambda_{j,g}-\lambda_{j,g}^{\geqslant\sigma})(P_{g,\theta}^{\geqslant
\sigma}\otimes I)\widetilde{R}^\sigma_{g,\theta} \big \| = O(
\sigma^2 g^{-2} ).
\end{equation}
By pulling $e^{-\delta \langle x \rangle}$ from $\Psi_j,$ we have using (\ref{eq:EstB1Factor}) that
\begin{equation}
\label{eq:B2Est2}
\left | (\Psi_{j,\overline{\theta}},\widetilde{R}^\sigma_{g,\theta} (\lambda_{j,g}-\lambda_{j,g}^{\geqslant\sigma}) P_{g,\theta}^{\geqslant \sigma}\otimes I
\widetilde{R}^\sigma_{g,\theta} W^{\leqslant \sigma}_{g,\theta}\widetilde{R}^\sigma_{g,\theta}\Psi_{j,\theta}) \right |  = O( \sigma^{5/2} g^{-2}).
\end{equation}
Using (\ref{Wqed5}), we have
\begin{align}
 \widetilde{R}^\sigma_{g,\theta} W^{\leqslant \sigma}_{g,\theta}\widetilde{R}^\sigma_{g,\theta} W^{\leqslant \sigma}_{g,\theta}\widetilde{R}^\sigma_{g,\theta} =& \underbrace{ g \widetilde{R}^\sigma_{g,\theta} W^{\leqslant \sigma}_{g,\theta} x\cdot A^{\leqslant
\sigma}_\theta(x)\widetilde{R}^\sigma_{g,\theta}}_{III} - \underbrace{ g \widetilde{R}^\sigma_{g,\theta} W^{\leqslant \sigma}_{g,\theta}
\widetilde{R}^\sigma_{g,\theta}   x\cdot  A^{\leqslant
\sigma}_\theta (x)}_{IV} \nonumber \\&+\underbrace{ \widetilde{R}^\sigma_{g,\theta} W^{\leqslant \sigma}_{g,\theta}\widetilde{R}^\sigma_{g,\theta} I  \widetilde{R}^\sigma_{g,\theta}}_{V} + \underbrace{ \widetilde{R}^\sigma_{g,\theta} W^{\leqslant \sigma}_{g,\theta}\widetilde{R}^\sigma_{g,\theta} II  \widetilde{R}^\sigma_{g,\theta}}_{VI}.\label{eq:B2Factor}
\end{align}
It follows from (\ref{StandardEst1}) and
(\ref{RtildeEst})-(\ref{eq:expdelta_tildeR}) and from
\eqref{StandardEst2} that
\begin{equation}
\label{eq:B2Est3} \left \| e^{-\delta \langle x \rangle } III \right
\| = O(\sigma)\ \mbox{and}\ \left \| e^{-\delta \langle x \rangle }
IV \right \| = O(\sigma + \sigma^{3/2}g^{-1} ).
\end{equation}
Since the operator $I$ is quadratic in $A_\theta^{ \leqslant \sigma }$, we obtain by putting it to the normal form and using again (\ref{StandardEst1}) and (\ref{RtildeEst})-(\ref{eq:expdelta_tildeR})
\begin{equation}
\label{eq:B2Est5}
\left \| e^{-\delta \langle x \rangle } V \right \| = O(\sigma^{3/2}).
\end{equation}
Finally, as above the fact than $II$ has better infrared behavior
than $A_\theta^{ \leqslant \sigma }$ by the factor $\omega$ leads to
\begin{equation}
\label{eq:B2Est6}
\left \| e^{-\delta \langle x \rangle } VI \right \| = O(\sigma^{2} g^{-2}).
\end{equation}
%
%
By pulling $e^{-\delta \langle x \rangle}$ from $\Psi_j$ and using
(\ref{eq:B2Factor})-(\ref{eq:B2Est6}) we find that
\begin{equation}
\label{eq:B2Est5}
( \Psi_{j,\overline{\theta}}, \widetilde{R}^\sigma_{g,\theta} W^{\leqslant \sigma}_{g,\theta}\widetilde{R}^\sigma_{g,\theta} W^{\leqslant \sigma}_{g,\theta}\widetilde{R}^\sigma_{g,\theta} \Psi_{j,\theta}) = O(\sigma^2 g^{-2}).
\end{equation}
Together with a factor of $\sigma$ obtained from the $z$
integration, (\ref{eq:B2Est1}), (\ref{eq:B2Est2}) and
(\ref{eq:B2Est5}) yield the estimate $B^2 = O(\sigma^3 g^{-2}).$

Instead of (\ref{eq:MetaStab}), Section 3, we have in the case of
non-relativistic QED
\begin{equation*}
\left ( \Psi_j , e^{-it H_g} f(H_g) \Psi_j \right ) =
e^{-it\lambda_{j,g}} + O(g^2\sigma^{-1})  + O(\sigma^2 g^{-2} ),
\ \ t\ge 0.
\end{equation*}
Optimizing and removing the $f$ dependence as in the proof of
Theorem \ref{main_theorem} gives
\begin{equation*}
\left ( \Psi_j , e^{-it H_g} \Psi_j \right ) = e^{-it\lambda_{j,g}}
+ O(g^{2/3}), \ \ t\ge 0.
\end{equation*}

\section{Proof of Theorem \ref{thm:resonpoles}}\label{sec:polesproof}

Let $ P_{\Omega}$ be the projection on the vacuum $\Omega$ in
${\mathcal F}_s$. We prove Theorem \ref{thm:resonpoles} for the set
${\mathcal D}'$ chosen explicitly as $${\mathcal D}':= \{\psi \in
{\mathcal D} | \ \ \| d \Gamma ( \omega^{-1/2}) (1 - P_{\Omega})
\psi \| < \infty \}$$
 for the Nelson model and as $${\mathcal D}':=
\{\psi \in {\mathcal D} | \ \ \|e^{\delta \langle x \rangle} d
\Gamma ( \omega^{-1/2}) (1 - P_{\Omega}) \psi \| < \infty\ \mbox{for
some}\ \delta >0 \}$$ for the QED one. Since $\mathcal{U}_\theta d
\Gamma ( \omega^{-1/2})= e^{\theta/2} d \Gamma (
\omega^{-1/2})\mathcal{U}_\theta$, the set ${\mathcal D}'$ is dense
in ${\mathcal D}$.

We conduct the proof for the Nelson model only. To extend it to the
QED one uses the methodology of Section \ref {sec:QEDproof}. As in
Sections \ref{sec:Hamiltonians} - \ref{sec:Ham}, the symbol
$H_{g,\theta}$ stands for the dilatation transformation, \eqref{I.10}
of the Nelson Hamiltonian $H_g = H_g^{N}$.

The RG analysis \cite{BFS1, BFS2}
shows that given $\delta>0$, there exist $g_{*} >0$ and $\varphi_{*}
\in (0, \varphi_0)$ s.t. for $g \le g_{*}$ and $\Im \theta \in
(\varphi_{*},\varphi_{0}) $, the spectrum of the operator
$H_{g,\theta}$ in the half-plane $ \{\Re z \le \Sigma -\delta\} $
lies in the union of wedges
\begin{equation*}
S_{j} := \lambda_{j,g} + \{z\in {\mathbb C} |\ \  |\arg(z) - \Im \theta|
\le \epsilon \},
\end{equation*}
where $\lambda_{j,g} = \lambda_j + O(g^2)$, $\Im\lambda_{j,g} \le 0$
and $\epsilon< |\Im\theta|$ is a positive number \footnote{The proof
for the QED model without the confinement assumption is given in
\cite{Sigal}. }. Moreover, the apices, $\lambda_{j,g}$, of these
wedges are the eigenvalues of $H_{g,\theta}$. If, in addition,
condition (C) holds for $\lambda_{j}$ then $\Im\lambda_{j,g} \le
-\mathrm{const}. ~ g^2.$


We take $z \in W_{\lambda_{j,g}}^{\varphi_1, \varphi_2}$ with
$\varphi_1 = \pi/2- \varphi_{0}$ and $\varphi_2 > 3\pi/2 -
\varphi_{*}$.
We want to estimate $( \psi , (H_{g,\theta} - z)^{-1} \psi ).$ Using an infrared cut-off as in section
\ref{sec:Hamiltonians}, we decompose
\begin{equation} \label{6.2}
H_{g,\theta} = \widetilde{H}_{g,\theta}^\sigma +
\widetilde{W}_{g,\theta}^{ \leqslant \sigma } ,
\end{equation}
see \eqref{2.16}. The infrared cut-off Hamiltonian
$\widetilde{H}_{g,\theta}^\sigma$ has an eigenvalue at
$\lambda_{j,g}.$ We use the second resolvent equation
\begin{equation}
\label{eq:ResolventId} (H_{g,\theta} -z)^{-1} =
(\widetilde{H}_{g,\theta}^\sigma -z)^{-1} +
(\widetilde{H}_{g,\theta}^\sigma -z)^{-1} \widetilde{W}_{g,\theta}^{
\leqslant \sigma } (H_{g,\theta} -z)^{-1} .
\end{equation}
Let $\widetilde{R}^{\sigma}_{g,\theta} (z):=
(\widetilde{H}_{g,\theta}^\sigma -z)^{-1}$ and let
$P^{\leqslant\sigma}_\Omega$ be the projection onto the vacuum state
of ${\mathcal F}_{s}^{\leqslant\sigma}$ and $\overline{P} = 1- P.$
Then
\begin{equation}
\label{eq:ResolventId2}
\widetilde{R}^{\sigma}_{g,\theta} (z)= \frac{
P^{\geqslant\sigma}_{g,\theta} \otimes P^{\leqslant\sigma}_\Omega
}{\lambda_{j,g} -z}  + \frac{ P^{\geqslant\sigma}_{g,\theta} \otimes
\overline{P}^{\leqslant\sigma}_\Omega}{\lambda_{j,g} + e^{-\theta}
H_f^{\leqslant\sigma} - z} + \widehat{R}_{g,\theta}^\sigma(z),
\end{equation}
where, as above,
\begin{equation}
\label{Rhat_decomp}
\widehat{R}_{g,\theta}^\sigma(z):=(\overline{P}^{\geqslant\sigma}_{g,\theta}\otimes
I) \widetilde{R}_{g,\theta}^\sigma(z).
\end{equation}

By our condition on $z$ we can pick $\theta$ so that
\begin{equation}\label{eq:choice_z}
\Re(e^{\theta}(\lambda_{j,g} - z)) \ge 0,
\end{equation}
i.e. $|\Im \theta +
\arg(\lambda_{j,g} - z)| \le \pi/2$. Then
\begin{equation}
\label{eq:2ndTerm} |( \psi, \frac{
P^{\geqslant\sigma}_{g,\theta} \otimes
\overline{P}^{\leqslant\sigma}_\Omega}{\lambda_{j,g} + e^{-\theta}
H_f^{\leqslant\sigma} - z}\psi )| \le
\|(H_f^{\leqslant\sigma})^{-1/2}\overline{P}^{\leqslant\sigma}_\Omega
\psi\|^2.
\end{equation}
(More generally, the l.h.s. is bounded by $|\lambda_{j,g} -
z|^{-\alpha}
\|(H_f^{\leqslant\sigma})^{-(1-\alpha)/2}\overline{P}^{\leqslant\sigma}_\Omega
\psi\|^2$ for $0 \le \alpha \le 1$.) Furthermore, an elementary
analysis of the $n-$photon sectors shows that
\begin{equation}\label{eq:D' ineq}
\|(H_f^{\leqslant\sigma})^{-1/2}\overline{P}^{\leqslant\sigma}_\Omega
\psi\| \le \| d \Gamma ( \omega^{-1/2}) \overline{P}_{\Omega} \psi
\|.
\end{equation}
Hence, by the definition of ${\mathcal D}',$ we have that, for all $\psi
\in {\mathcal D}',$
\begin{equation}\label{eq:2ndTermFinal} |( \psi, \frac{
P^{\geqslant\sigma}_{g,\theta} \otimes
\overline{P}^{\leqslant\sigma}_\Omega}{\lambda_{j,g} + e^{-\theta}
H_f^{\leqslant\sigma} - z}\psi )| \le \mathrm{C} .
\end{equation}

\begin{figure}[htbp]
\centering
\resizebox{0.8\textwidth}{!}{ \includegraphics{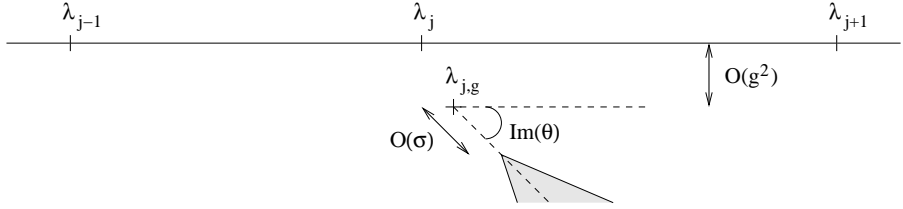} }
\caption{\textbf{Spectrum of $\widetilde{H}_{g,\theta}^{ \geqslant \sigma }$ near $\lambda_{j,g}$} } \label{spectrum_Hgesigma_2}
\end{figure}

Next, to estimate $\widehat{R}_{g,\theta}^\sigma(z)$, see  Eq.
\eqref{Rhat_decomp}, we use the representation \eqref{eq:rep_rhat}.
Applying to $H_{g,\theta}^{ \geqslant \sigma }$ a renormalization
group analysis as in \cite{BFS1, BFS2, BCFS, FGS}, one can show that
the spectrum of $\widetilde{H}_{g,\theta}^{ \geqslant \sigma }$ is
of the form pictured in Figure \ref{spectrum_Hgesigma_2}, and that
for $|z-\lambda_{j,g}|\le \sigma/2$ and $\omega \ge 0$
\begin{equation}
\label{eq:RegTermHat} \| \widehat{R}_{g,\theta}^{ \geqslant \sigma
}(z-e^{-\theta}\omega)\| \le \mathrm{C}(\sigma +\omega)^{-1},
\end{equation}
which, together with \eqref{eq:rep_rhat}, implies,  for
$|z-\lambda_{j,g}|\le \sigma/2$ and $n=0, 1/2, 1,$ the estimate
\begin{equation}
\label{eq:RegTerm} \| (H_{f}^{ \leqslant \sigma })^{n}
\widehat{R}_{g,\theta}^\sigma(z)\| \le \mathrm{C}\sigma^{n-1},
\end{equation}
for some constant C.

\noindent Eqns \eqref{eq:ResolventId2},
\eqref{eq:2ndTermFinal} and \eqref{eq:RegTerm} imply that, for $ \psi
\in {\mathcal D}',$
\begin{equation}
\label{eq:RtildeEstim} | ( \psi, (\widetilde{R}^{\sigma}_{g,\theta}
(z)- \frac{ P^{\geqslant\sigma}_{g,\theta} \otimes
P^{\leqslant\sigma}_\Omega }{\lambda_{j,g} -z})\psi ) | \le
\mathrm{C}/\sigma,
\end{equation}

Finally we estimate the last term on the r.h.s. Eq.
\eqref{eq:ResolventId}. Recall that
\begin{equation}
\label{eq:tildeWa}
\widetilde{W}_{g,\theta}^{\leqslant\sigma}=W_{g,\theta}^{ \leqslant
\sigma } - \left ( \lambda_{j,g} - \lambda_{j,g}^{ \geqslant \sigma
} \right ) \mathcal{V}^{-1}\big( P_{g,\theta}^{ \geqslant \sigma }
\otimes I \big) \mathcal{V} ,
\end{equation}
where
\begin{equation}
W_{g,\theta}^{ \leqslant \sigma } := W_{g,\theta} - W_{g,\theta}^{
\geqslant \sigma } = g \phi ( G_{x,\theta}^{ \leqslant \sigma } ).
\end{equation}
Below, we let $\sigma \rightarrow 0,$ as $|\lambda_{j,g} -z|
\rightarrow 0$. Hence we have to estimate $\lambda_{j,g} -
\lambda_{j,g}^{ \geqslant \sigma }$ for any $\sigma > 0$. We claim
that
\begin{equation} \label{lambda difference} |\lambda_{j,g} -
\lambda_{j,g}^{ \geqslant \sigma }| =O \left((g
\sigma^{1/2+\mu})^{2} \right).
\end{equation}
This estimate is proven in the proposition at the end of this
section.

Iterating the last term on the r.h.s. of Eq. \eqref{eq:ResolventId} we
see that the worst term is $\widetilde{R}^{\sigma}_{g,\theta}(z)
\widetilde{W}_{g,\theta}^{ \leqslant \sigma }
\widetilde{R}^{\sigma}_{g,\theta}(z)$. We use the decomposition
\eqref{eq:ResolventId2}. Since the operator $W_{g,\theta}^{
\leqslant \sigma }$ is in normal form, we see that the term coming
from sandwiching it between the first term on the r.h.s. of
\eqref{eq:ResolventId2} vanishes. Thus, it remains to consider the
terms
\begin{equation} \label{part1}
\widetilde{R}^{\sigma}_{g,\theta}(z)
 \left ( \lambda_{j,g} - \lambda_{j,g}^{ \geqslant \sigma
} \right ) \big( P_{g,\theta}^{ \geqslant \sigma } \otimes I \big)
\widetilde{R}^{\sigma}_{g,\theta}(z)  ,\end{equation}
$$[\frac{ P^{\geqslant\sigma}_{g,\theta} \otimes
P^{\leqslant\sigma}_\Omega }{\lambda_{j,g} -z} + \frac{
P^{\geqslant\sigma}_{g,\theta} \otimes
\overline{P}^{\leqslant\sigma}_\Omega}{\lambda_{j,g} + e^{-\theta}
H_f^{\leqslant\sigma} - z} + \widehat{R}_{g,\theta}^\sigma(z)]$$
\begin{equation}\label{part2}
\times W_{g,\theta}^{ \leqslant \sigma }[ \frac{
P^{\geqslant\sigma}_{g,\theta} \otimes
\overline{P}^{\leqslant\sigma}_\Omega}{\lambda_{j,g} + e^{-\theta}
H_f^{\leqslant\sigma} - z} + \widehat{R}_{g,\theta}^\sigma(z)]\end{equation}
and the term obtained by switching the right and left factors in (6.17).

We note that, by the decomposition
\eqref{eq:ResolventId2} and the definition of
$\widehat{R}_{g,\theta}^\sigma(z)$, Eq. \eqref{part1} can be written as
\begin{equation}
\left ( \lambda_{j,g} - \lambda_{j,g}^{
\geqslant \sigma } \right )
 \frac{ P^{\geqslant\sigma}_{g,\theta} \otimes
P^{\leqslant\sigma}_\Omega }{(\lambda_{j,g} -z)^2} + \left ( \lambda_{j,g} - \lambda_{j,g}^{ \geqslant \sigma } \right
)
 \frac{ P^{\geqslant\sigma}_{g,\theta} \otimes
\overline{P}^{\leqslant\sigma}_\Omega }{(\lambda_{j,g} + e^{-\theta}
H_f^{\leqslant\sigma} - z)^2}.\end{equation} Using \eqref{lambda
difference} we obtain the following estimate for \eqref{part1}:
\begin{equation}\label{estpart1}
\eqref{part1} = O \left((g \sigma^{1/2+\mu}|\lambda_{j,g}
-z|^{-1})^{2} \right).
\end{equation}

To estimate \eqref{part2}, we first observe that, due to \eqref{eq:choice_z}, we have that,
for $n=0,1/2,1$
\begin{equation}
\label{part2a} \| (H_f^{\leqslant\sigma})^n (\lambda_{j,g} +
e^{-\theta} H_f^{\leqslant\sigma} - z)^{-1}\| \le
\mathrm{C}|\lambda_{j,g}  - z|^{n-1}.
\end{equation}
Assume $\sigma \ge |z-\lambda_{j,g} |$. Using estimates
\eqref{estimate_Wgleq}, \eqref{eq:RegTerm}  and \eqref{part2a} (or
\eqref{part2a1}), the fact that $P^{\leqslant\sigma}_\Omega
a^*(G_{x,\theta}^{ \leqslant \sigma }) =0 $ and standard estimates on
the creation and annihilation operators, and remembering the
condition that $\Re(e^{\theta}(\lambda_{j,g} - z)) \ge 0,$ we obtain the bound
\begin{equation}
 \label{estpart2} \|\widetilde{R}^{\sigma}_{g,\theta}(z)
W_{g,\theta}^{ \leqslant \sigma } \widetilde{R}^{\sigma}_{g,\theta}(z) \|\le
\mathrm{C} \frac{1}{|z-\lambda_{j,g}| }g\sigma^{\frac{1}{2}+\mu}
(\frac{1}{|z-\lambda_{j,g}|^{1/2}}+\frac{1}{\sigma^{1/2}}).
\end{equation}
This together with \eqref{estpart1} yields
\begin{equation}
 \label{eq:1st iteration} \|\widetilde{R}^{\sigma}_{g,\theta}(z)
\widetilde{W}_{g,\theta}^{ \leqslant \sigma }
\widetilde{R}^{\sigma}_{g,\theta}(z) \| \le \mathrm{C}
\frac{g\sigma^{\frac{1}{2}+\mu} }{|z-\lambda_{j,g}|
}(\frac{1}{\sigma^{1/2}} + \frac{1}{|z-\lambda_{j,g}|^{1/2}
}+\frac{g\sigma^{\frac{1}{2}+\mu} }{|z-\lambda_{j,g}| }) .
\end{equation}

Since, as we mentioned, the higher order iterates of
\eqref{eq:ResolventId} are estimated similarly and lead to improved
estimates, we conclude, assuming
$\sigma \ge |z-\lambda_{j,g} |$, that
\begin{equation}
 \label{eq:2nd term in 6.3} \|\widetilde{R}^{\sigma}_{g,\theta}(z)
\widetilde{W}_{g,\theta}^{ \leqslant \sigma } R_{g, \theta}(z) \|\le
\mathrm{C}( \frac{g\sigma^{\frac{1}{2}+\mu}
}{|z-\lambda_{j,g}|^{3/2} }+\frac{g^2\sigma^{1+2\mu}
}{|z-\lambda_{j,g}|^{2} }),
\end{equation}
where $R_{g, \theta}(z) :=(H_{g,\theta} -z)^{-1}$.

It follows from  \eqref{eq:ResolventId}, \eqref{eq:RtildeEstim} and
\eqref{eq:2nd term in 6.3} that, for $g$ small enough,
\begin{equation}
\label{eq:NormDiff} |( \psi, ((H_{g,\theta} -z)^{-1} -
\frac{1}{\lambda_{j,g}-z} P^{\geqslant\sigma}_{g,\theta} \otimes
P^{\leqslant\sigma}_\Omega)\psi ) | \le \mathrm{C}(\frac{1}{\sigma}
+ \frac{g\sigma^\alpha}{r^{3/2}} +
\frac{g^2\sigma^{2\alpha}}{r^{2}}),
\end{equation}
where $r:=|z-\lambda_{j,g}|$ and $\alpha:= 1/2 +\mu$, for some
constant C, provided
$\sigma \ge |z-\lambda_{j,g} |$. We now pick
$\sigma= r^{\beta}g^{-(3/2+ \mu)^{-1}},$ where
$\beta:=(1+\frac{2}{3}\mu)^{-1}$. By our assumption,
 $\beta<1$ and therefore $\sigma
> r .$ Then, for this choice of $\sigma ,$
\begin{equation*}
|( \psi, ((H_{g,\theta} -z)^{-1} - \frac{1}{\lambda_{j,g}-z}
P^{\geqslant\sigma}_{g,\theta} \otimes
P^{\leqslant\sigma}_\Omega)\psi) | \le \mathrm{C}g^{(3/2+
\mu)^{-1}}r^{-\beta}.
\end{equation*}
Let $\psi_{\theta}: = \mathcal{U}_{\theta} \psi$. The last estimate, together
with the relation
\begin{equation}
\label{eq:CombesRelation} ( \psi,(H_g-z)^{-1}\psi ) =
(
\psi_{\bar{\theta}},(H_{g,\theta}-z)^{-1}\psi_{\theta} ),
\end{equation}
implies (i) and (ii) in Theorem \ref{thm:resonpoles}, with $\beta=
(1+\frac{2}{3}\mu)^{-1}$. $\Box$

\begin{remarque}The expression for $\beta$ can be improved if one uses \eqref{eq:D' ineq}
to conclude that,  for $n=0,1/2,$
\begin{equation}\label{part2a1}
\| (H_f^{\leqslant\sigma})^n (\lambda_{j,g} + e^{-\theta}
H_f^{\leqslant\sigma} - z)^{-1}\psi\| \le \mathrm{C}|\lambda_{j,g} -
z|^{n-1/2} \| d \Gamma ( \omega^{-1/2}) \overline{P}_{\Omega} \psi
\|,
\end{equation}
which is better than \eqref{part2a}. This estimate leads to the
inequality
\begin{equation}
 \label{estpart2a} |( \psi, \widetilde{R}^{\sigma}_{g,\theta}(z)
W_{g,\theta}^{ \leqslant \sigma } \widetilde{R}^{\sigma}_{g,\theta}(z) \psi)
|\le \mathrm{C} \frac{1}{|z-\lambda_{j,g}|
}g\sigma^{\frac{1}{2}+\mu}
(\frac{\sigma^{1/2}}{|z-\lambda_{j,g}|^{1/2}}+\frac{1}{\sigma^{1/2}}),
\end{equation}
which has a better r.h.s than \eqref{estpart2}.
\end{remarque}
\begin{remarque} \label{transformations} To define resonances
for the QED model it is technically more convenient to use a family
of unitary transformations different from the dilatation one (see
\cite{Sigal}).
\end{remarque}
\begin{prop}\label{lambda_diff_estimate}
Under the conditions of Theorem \ref{thm:resonpoles}, we have for
any $\sigma>0$
\begin{equation}
\lambda_{j,g} - \lambda_{j,g}^{ \geqslant \sigma } = O( g^2
\sigma^{1+\mu} ).
\end{equation}
\end{prop}
\begin{demo}. To prove \eqref{lambda difference} we use the RG approach. Here we only point out particularities of the present problem and outline the general
strategy; technical details can be found in \cite{BFS1, BFS2, BCFS,
FGS} (see also \cite{Sigal} for the QED case). Since we do not go
into details,
we use the Feshbach-Schur map, rather than the smooth Feshbach-Schur
map, to underpin our construction. The former (\cite{BFS1, BFS2}) is
simpler to formulate but the latter (\cite{BCFS, GH, FGS})  is easier to
handle technically. Our strategy follows (\cite{FGS}).

First we apply the Feshbach-Schur map  $\mathcal{F}_{P_{\rho_0}}$
associated to the projection $P_\rho :=
P^{\geqslant\sigma}_{g,\theta} \otimes
\chi^{\leqslant\sigma}_{\rho},$ where
$\chi^{\leqslant\sigma}_{\rho}:=
\chi_{H_f^{\leqslant\sigma}\leqslant\rho}$. For $z \in D(
\lambda_{j,g}^{ \geqslant \sigma} , \sigma/2 )$ and $\rho_0 = \sigma
$, the operator $H_{g,\theta} - z $ is in the domain of
$\mathcal{F}_{ P_{\rho_0} }$. Indeed, an easy estimate shows that
the operator $ \overline{P}_{\rho_0} H_{g,\theta}^{ \sigma }
\overline{P}_{\rho_0}  - z $ is invertible on $\mathrm{Ran}
\overline{P}_{\rho_0}$ and $\|[ H_f^{ \leqslant \sigma } + \sigma ]\overline{P}_{\rho_0} \left
[ \overline{P}_{\rho_0} H_{g,\theta}^{ \sigma }
\overline{P}_{\rho_0} - z \right ]^{-1}\overline{P}_{\rho_0} \| \le
\mathrm{C}$. Since $ \| [ H_f^{ \leqslant \sigma } + \sigma ]^{-1/2} W_{g,\theta}^{ \leqslant \sigma } [ H_f^{ \leqslant \sigma } + \sigma ]^{-1/2}
 \| \leq \mathrm{C} g \sigma^{\mu}$, we see by
Neumann series expansion that the operator $ \overline{P}_{\rho_0}
H_{g,\theta} \overline{P}_{\rho_0}  - z$ is invertible on $\mathrm{Ran}
\overline{P}_{\rho_0}$ and $\|\overline{P}_{\rho_0} \left [
\overline{P}_{\rho_0} H_{g,\theta} \overline{P}_{\rho_0}  - z \right
]^{-1}\overline{P}_{\rho_0} \| \le \mathrm{C}/\sigma$. Hence the operator
$H_{g,\theta} - z $ is in the domain of $\mathcal{F}_{ P_{\rho_0}
}$, as claimed. Next, we note that
$$\mathcal{F}_{ P_{\rho_0}  }( H_{g,\theta} - z ) =P^{\geqslant\sigma}_{g,\theta}
\otimes H_z,$$
where the operator $H_z$ acts on $\mathrm{Ran}
 \chi^{\leqslant\sigma}_{\rho_0} \subset \mathcal{F}_s^{ \leqslant \sigma }$ and is given by
\begin{equation}\label{eq_Feshbach1}
H_z := \chi^{\leqslant\sigma}_{\rho_0}(
\psi^{\geqslant\sigma}_{g,\theta},  \big( \lambda_{j,g}^{ \geqslant
\sigma}- z + H_f^{\leqslant\sigma}+  W_{g,\theta}^{ \leqslant
\sigma} + U \big)\psi^{\geqslant\sigma}_{g,\theta} )
\chi^{\leqslant\sigma}_{\rho_0} ,
\end{equation}
where $U:= -W_{g,\theta}^{ \leqslant \sigma } \overline{P}_{\rho_0}
\left [ \overline{P}_{\rho_0} H_{g,\theta} \overline{P}_{\rho_0}  -
z \right ]^{-1} \overline{P}_{\rho_0} W_{g,\theta}^{ \leqslant
\sigma } $.

By the isospectrality of the Feshbach-Schur map (see \cite{BFS1,
BFS2, BFS3, BCFS, FGS}), we have that $z\in D( \lambda_{j,g}^{
\geqslant \sigma} , \sigma/2 )$ is an eigenvalue of  $H_{g,\theta}$
iff $0$ is an eigenvalue of $H_z$. To investigate the spectral
properties of $H_z,$ we make use of the renormalization group method.

As a first step, we rewrite the operator $H_z$  in a generalized
normal form. To this end we  expand the resolvent on the r.h.s. in
a Neumann series in $W_{g,\theta}^{ \leqslant \sigma }$ and normal
order the creation and annihilation operators not entering the
expression for $H_f^{\leqslant\sigma}$.
This brings the operator $H_z$ to the form (see \cite{BFS1, BFS2,
BCFS, FGS})
\begin{equation}\label{decompHz}
H_z :=  \chi^{\leqslant\sigma}_{\rho_0}(E_z +T_z+W_z)
\chi^{\leqslant\sigma}_{\rho_0},
\end{equation}
where $ E_z$ is a number (more precisely, a complex function of $z$
and other parameters), $T_z$ is a differentiable function of
$H_f^{\leqslant\sigma}$ and $W_z$ is an operator in the generalized
normal form that is a sum of terms with at least one creation or
annihilation operator. A standard computation gives that $ E_z:=
\lambda_{j,g}^{ \geqslant \sigma}- z + \Delta E_z$, with
$$ \Delta
E_z:= - \int ( \psi^{\geqslant\sigma}_{g,\theta},
G_{x,\theta}^{ \leqslant \sigma }(k)
\overline{P}^{\geqslant\sigma}_{g,\theta}  \left [
\overline{P}^{\geqslant\sigma}_{g,\theta} H_{g,\theta}^{ \geqslant
\sigma } \overline{P}^{\geqslant\sigma}_{g,\theta} + e^{-\theta}
\omega - z \right ]^{-1} \overline{P}^{\geqslant\sigma}_{g,\theta}
G_{x,\theta}^{ \leqslant \sigma
}(k)\psi^{\geqslant\sigma}_{g,\theta} ) dk + h.o.t.,$$
$$T_z:= H_f^{\leqslant\sigma} - \int
( \psi^{\geqslant\sigma}_{g,\theta}, G_{x,\theta}^{ \leqslant \sigma
}(k) f(H_f^{\leqslant\sigma}+\omega) G_{x,\theta}^{ \leqslant \sigma
}(k)\psi^{\geqslant\sigma}_{g,\theta} ) dk + h.o.t.,$$
$$W_z:=   (
\psi^{\geqslant\sigma}_{g,\theta}, \big(W_{g,\theta}^{\leqslant
\sigma }- \int \int G_{x,\theta}^{ \leqslant \sigma }(k) a^*(k)
f(H_f^{\leqslant\sigma}+\omega +\omega') a(k') G_{x,\theta}^{
\leqslant \sigma }(k') dk dk'\big)\psi^{\geqslant\sigma}_{g,\theta}
) + h.o.t. ,$$
where $f(H_f^{\leqslant\sigma}):= \overline{P}_{\rho_0}  \left [
\overline{P}_{\rho_0} (H_{g,\theta}^{ \geqslant \sigma } +
e^{-\theta} H_f^{\leqslant\sigma}) \overline{P}_{\rho_0} - z \right
]^{-1} \overline{P}_{\rho_0} $. Clearly,
\begin{equation} \label{estimHz}
\Delta E_z=O \left((g \sigma^{1/2+\mu})^{2} \right)\ \textrm{and}\
\chi^{\leqslant\sigma}_{\rho_0}W_z \chi^{\leqslant\sigma}_{\rho_0}
=O \left(g \sigma^{1+\mu} \right).
\end{equation}
%

Let $a^\#(k)$ stand for either $a(k)$ or $a^*(k)$, $k \in
\mathbb{R}^3$. We define the \textit{scaling transformation}
$S_\rho: \mathcal{B}[\mathcal{F}_s^{ \leqslant \sigma }] \to
\mathcal{B}[\mathcal{F}_s^{ \leqslant \sigma/\rho }]$, by
%
%
%
\begin{equation} \label{IV.5}
S_\rho(I) := I, \quad  S_\rho (a^\#(k)) := \ \rho^{-3/2} \, a^\#(
\rho^{-1} k),
\end{equation}
%
and the dilatation transform, by
$A_\rho(A) \ := \ \rho^{-1}A$. Now we rescale the operator $H_z$ as
$H^{(0)}_z := A_\sigma(S_\sigma(H_z))$. The new operator acts on
$\mathrm{Ran} \chi^{\leqslant 1}_{1} \subset \mathcal{F}_s^{
\leqslant 1}$. The last estimate in \eqref{estimHz} and an estimate
on the derivative of $T_z$ as a function of $H_f^{\leqslant\sigma}$,
which we do not display here, show that the operator $H^{(0)}_z$ is
in the domain of the Feshbach-Schur map $\mathcal{F}_{
\chi^{\leqslant\sigma}_{\rho} }$,
provided $1/2 \ge \rho \gg g \sigma^{\mu}$ and $\rho \gg |E_z|
/\sigma$ (the latter inequality is also considered as a restriction
on $z$).

If we neglect the term $W_z$ in $H^{(0)}_z $ (see \eqref{decompHz})
then the remaining operator has the vacuum $\Omega$ as an eigenvector corresponding to the eigenvalue $0$, provided $z$ solves the equation $E^{(0)}_z:=(
\Omega,H^{(0)}_z\Omega ) = E_z /\sigma =0$. One can show
(\cite{FGS})  that this equation has a unique solution
$\lambda_{j,g}^{ (1)} = \lambda_{j,g}^{ \geqslant \sigma} +O
\left((g \sigma^{1/2+\mu})^{2} \right)$. By the isospectrality
mentioned above, this is our first approximation to $\lambda_{j,g}$.

Now we introduce the decimation map $F_\rho: = \mathcal{F}_{
\chi^{\leqslant\sigma}_{\rho} }.$
%
On the domain of the decimation map $F_\rho$ we define the
renormalization map $\mathcal{R}_\rho$ as \footnote{In principle,
the rescaling is not needed for the argument that follows, but we
use it, since it is used the machinery developed in \cite{BFS1,
BFS2, BCFS, FGS} and used here. }
\begin{equation} \label{IV.6}
\mathcal{R}_\rho:=  A_\rho\circ S_\rho\circ F_\rho.
\end{equation}
By the above, the operator $H^{(0)}_z$ is in the domain of the
decimation map $F_{ \rho }$ and therefore in the domain the
renormalization map $\mathcal{R}_\rho$, provided $1/2 \ge \rho \gg g
\sigma^{\mu}$ and $\rho \gg |E_z| /\sigma$.  Iterating this map as
in \cite{FGS} we obtain a sequence of operators $H^{(n)}_z,\ n=0, 1,
2, ..., $ (Hamiltonians on scales $0, 1, ...$) acting on the space
$\mathrm{Ran} \chi^{\leqslant 1}_{1} \subset \mathcal{F}_s^{
\leqslant 1}$. Again, one argues that $0$ is an approximate
eigenvalue of the operators $H^{(n)}_z,$ provided $z$ satisfies the
equations $E^{(n)}_z:=( \Omega,H^{(n)}_z\Omega) =0$. Namely, one
proves that the equations  $( \Omega,H^{(n)}_z\Omega) =0$ in $z$
have have unique solutions $\lambda_{j,g}^{ (n)}$ satisfying
$$\lambda_{j,g}^{ (n)} = \lambda_{j,g}^{ \geqslant \sigma} +O
\left((g \sigma^{1/2+\mu})^{2} \right)$$ and $|\lambda_{j,g}^{ (n)}-
\lambda_{j,g}^{ (n-1)}| \le const\ \rho^n$ (see  \cite{FGS},
Proposition V.3). Consequently, $\lambda_{j,g}^{ (n)}$  converge,
$\lambda_{j,g}^{ (n)} \rightarrow \lambda_{j,g},$ as $n \rightarrow
\infty $. By the isospectrality of $\mathcal{R}_{ \rho }$
we conclude
that the operator $H^{(0)}_z$ has a simple eigenvalue $0,$ provided
$z=\lambda_{j,g}$ (see  \cite{FGS}, Theorem V.2).
Hence, by the isospectrality of the Feshbach-Schur map, the operator
$H_{g,\theta}$ has a unique eigenvalue $\lambda_{j,g}$ in the disc
$D( \lambda_{j,g}^{ \geqslant \sigma} , \sigma/2 )$ and this
eigenvalue satisfies \eqref{lambda difference}. Since on the other
hand $\lambda_{j,g}^{ \geqslant \sigma }= \lambda_j +O(g^2)$ is the
unique eigenvalue of the operator $H_{g,\theta}^{ \geqslant \sigma
}$ bifurcating from the eigenvalue $\lambda_j$ of $H_0$, we conclude
that $\lambda_{j,g}$ is the unique eigenvalue of the operator
$H_{g,\theta}$ emerging from the eigenvalue $\lambda_j$ of $H_0$.
\end{demo}

\footnotesize{\normalsize}\par

\vspace{0.5cm}

\noindent $^a$ Department of Mathematics, University of Toronto, Toronto, Ontario, Canada M5S 2E4. \\ E-mail: walid@math.utoronto.ca; im.sigal@utoronto.ca.

\noindent $^b$ Laboratoire de Math{\'e}matiques EDPPM, UMR-CNRS 6056, Universit{\'e} de Reims, Moulin de la Housse - BP 1039, 51687 REIMS Cedex 2, France. \emph{Current address}: Institut for Matematiske Fag, Aarhus Universitet, Ny Munkegade, 8000 Aarhus C, Denmark. E-mail: faupin@imf.au.dk.

\noindent $^c$ Institute for Theoretical Physics, ETH Zurich, CH-8093, Switzerland. E-mail: juerg@phys.ethz.ch.

\noindent $\#$ The research of these authors is supported by NSERC
under Grant $\#$ 7901.

\noindent $*$ Supported by the Centre for Theory in Natural Science

\end{document}